\documentclass[letterpaper,twocolumn,10pt]{article}

\usepackage[frozencache=true,cachedir=minted-cache]{minted} 
\usepackage{hyperref}
\hypersetup{colorlinks,allcolors=black}
\usepackage{usenix}

\usepackage{xcolor,colortbl}
\usepackage[english]{babel}
\usepackage{blindtext}
\usepackage{multicol}

\usepackage{algorithm}
\usepackage{algorithmicx}
\usepackage{algpseudocode}
\usepackage{amsmath}

\usepackage{tikz}
\usepackage{amsmath}
\usepackage{xspace}
\usepackage{booktabs}
\usepackage{subcaption}

\usepackage{multirow}

\usepackage{siunitx}
\usepackage{pifont}

\def\Snospace~{\S{}}

\usepackage{enumitem}

\usepackage{url}

\usepackage[small, compact]{titlesec}
\usepackage[font=small]{caption}

\newcommand{\CR}[1]{{#1}}
\newcommand{\autorefsuffix}[2]{\hyperref[#1]{\autoref*{#1}#2}}

\algnewcommand{\IfThenElse}[3]{
  \State \algorithmicif\ #1\ \algorithmicthen\ #2\ \algorithmicelse\ #3}
  
\begin{document}

\date{}

\newcommand{\sys}{Collie\xspace}

\renewcommand{\algorithmicrequire}{\textbf{Input:}}  
\renewcommand{\algorithmicensure}{\textbf{Output:}}

\title{\sys: Finding Performance Anomalies in RDMA Subsystems}

\author{
\rm{Xinhao Kong$^{\text{1,2}}$ \enskip
    Yibo Zhu$^{\text{2}}$ \enskip
    Huaping Zhou$^{\text{2}}$ \enskip
    Zhuo Jiang$^{\text{2}}$ \enskip
} \\
\rm{
    Jianxi Ye$^{\text{2}}$ \enskip
    Chuanxiong Guo$^{\text{2}}$ \enskip
    Danyang Zhuo$^{\text{1}}$ \enskip}\\
\\
  {$^\text{1}$Duke University\enskip $^\text{2}$ByteDance Inc.}
} 
\maketitle

\section*{Abstract}
High-speed RDMA networks are getting rapidly adopted in the industry for their low latency and reduced CPU overheads. To verify that RDMA can be used in production, system administrators need to understand the set of application workloads that can potentially trigger abnormal performance behaviors (e.g., unexpected low throughput, PFC pause frame storm). We design and implement \sys, a tool for users to systematically uncover performance anomalies in RDMA subsystems without the need to access hardware internal designs. Instead of individually testing each hardware device (e.g., NIC, memory, PCIe), \sys is holistic, constructing a comprehensive search space for application workloads. \sys then uses simulated annealing to drive RDMA-related performance and diagnostic counters to extreme value regions to find workloads that can trigger performance anomalies. We evaluate \sys on combinations of various RDMA NIC, CPU, and other hardware components. \sys found 15 new performance anomalies. All of them are acknowledged by the hardware vendors. 7 of them are already fixed after we reported them. We also present our experience in using \sys to avoid performance anomalies for an RDMA RPC library and an RDMA distributed machine learning framework.

\section{Introduction}

Data center applications relentlessly demand low packet latency and high CPU efficiency. That makes Remote Direct Memory Access (RDMA) an appealing solution for cloud providers and other data center operators. Today, many top companies have already adopted RDMA in their data centers~\cite{zhu2015dcqcn, li2019hpcc, guo2016rdma}. RDMA has been integrated into many application domains, such as graph processing~\cite{buragohain2020a1, shi2016graph}, data stores~\cite{aleksandar2014farm, kalia2014herd}, and deep learning~\cite{xue2019rdma, jiang2020byteps}. 

To deploy RDMA in production, i.e., using RoCEv2 for Ethernet-based data center network, we need to make sure that the RDMA network performance can meet our expectations, free of performance anomalies like low throughput and pause frame storm~\cite{guo2016rdma, zhu2015dcqcn, storm2019systor, hu2016pfc}. This is important because applications require high-performance RDMA networks to deliver their service-level objectives (SLO). Furthermore, some abnormal behaviors, like pause frame storms, can cause catastrophic consequences including deadlocking the entire data center network ~\cite{guo2016rdma, hu2016pfc, yixiao2021nsdi, gfc2019sigcomm}.

We have encountered the following anomalies in our RoCEv2 production environment:
\begin{itemize}
    \item A particular application workload's performance of the same RDMA NIC (RNIC) varies substantially on servers with only a slight difference in their PCIe specifications. 
    \item A specific application workload only triggers pause frame storms with certain NUMA settings on a particular RNIC combined with particular server hardware. 
    \item A particular application workload triggers pause frame storms with only a single connection on a particular RNIC from a particular vendor. 
\end{itemize}

Although we collaborate with the most reliable vendors and they have conducted extensive tests on individual devices, the entire RDMA subsystem still has anomalies. The RDMA subsystem consists of RNICs and other server hardware that interacts with the RNICs. Our observation is that most of the anomalies are highly related to the interactions between RNICs and rest of the server hardware. Additional integration tests are thus critical, and we usually conduct these tests on our own because of two reasons. First, vendors cannot access our highly customized hardware, system configurations, and applications. Second, anomalies are too critical for the reliability and performance of the entire data center network, and we cannot completely rely on third parties for testing.

Currently, there are two approaches to conduct tests over the entire subsystem. The first approach is to run simple test benchmarks (e.g., \texttt{Perftest}~\cite{perftest}) to conduct basic throughput and latency tests. The second approach is to run a set of representative RDMA applications. Unfortunately, these two approaches are not able to comprehensively uncover RDMA subsystem anomalies. The fundamental problem is that these approaches only test simple or existing workloads. They therefore fail to capture anomalies comprehensively because real application workloads change over time.
In addition, even if an anomaly is found with an application workload, application developers do not know how to modify the workload to avoid the anomaly.

Our goal for this paper is to explore the possibility of \textbf{systematic search} for application workloads that can trigger performance anomalies in RDMA subsystems. Finding these anomalies for the vendors can help them improve their hardware and thus improve the reliability and the performance of the entire data center network. Besides, the systematic approach can help developers understand the conditions to trigger such anomalies and how to avoid them by changing application workloads. 

To realize this goal, the first question is \textit{how to formally define an anomaly?} Having such a definition is difficult because application performance highly depends on the workload and the hardware. In this paper, we focus on two types of performance anomalies that can be precisely defined: no PFC pause frames if the network is not congested and throughput should be bottlenecked either by bits/second or packets/second as in RNIC specification.

Given this definition, we still need to address three challenges. The first challenge is how to build a comprehensive workload search space. An ideal approach for testing with the entire RDMA subsystem is to exactly modeling each component and then construct the search space. However, this is extremely hard for us, given the black-box nature of RNIC and other hardware components. The second challenge is even after we successfully construct a comprehensive enough search space, how can we search efficiently? The search space is inherently very large because RDMA subsystems are complicated. For example, traffics within an RDMA subsystem can be from/to different memory devices (e.g., main memory and GPU memory) and the transportation setting for a given workload is various (e.g., number and type of connections). Conducting tests blindly in such a large space is inefficient. The third challenge is how to find the complicated triggering conditions of such anomalies? This is important both during the search and after the search. During the search, we need the triggering condition to avoid testing similar application workloads for the same anomaly to speed up the search. After the search, we need to use these conditions to help developers avoid anomalies.

To this end, we design and implement \sys, the first tool to systematically uncover RDMA subsystem performance anomalies, with the following three ideas.

Our first idea is to construct the search space from a developer's perspective. Though the underlying hardware is various and opaque to us, the narrow-waist RDMA programming abstractions (i.e., \textit{verbs}) are clearly defined and stable. All application workloads can be interpreted as a combination of \textit{verbs} operations. We carefully analyze the standard \textit{verbs} library and the design decisions developers are allowed to make (the request pattern, how RDMA buffers are allocated, etc.). Moreover, to cover the entire RDMA subsystem, we analyze all the potential data flows within a given server configuration. In this way, \sys constructs a comprehensive search space for application workloads in the domain of RDMA subsystem, including the host of the network traffic (e.g., GPU connected to a different PCIe bridge from the RNIC, DRAM from a different CPU socket), message sizes, number of connections, and memory region configurations. 

Our second idea is that we can use two sets of counters to guide the search. The first set is the performance counters (e.g., bits per second), which are provided by all commodity RNICs and other hardware components. In addition, modern commodity RNICs and other hardware components provide diagnostic counters (e.g., PCIe backpressure). Diagnostic counters are mapped to particular unexpected events that happen to the hardware components. These counters are currently only used for debugging and monitoring purposes.  \sys uses search algorithms based on simulated annealing to maximize/minimize counter values to uncover anomalies. 

Our third idea is to find the minimal area in the search space that covers the found anomalies. We call this area (i.e., the conditions to trigger the anomaly) the minimal feature set (MFS). \sys includes a MFS algorithm to test each feature that an anomaly has (e.g., number of connections) and generate the necessary conditions set. With the MFS algorithm, \sys can further improve search efficiency by avoiding redundant tests of the same area. Also, finding the triggering conditions of an anomaly allows developers to avoid the anomaly by breaking one of the provided conditions.

We evaluate \sys on 8 different RDMA subsystems, including 6 types of RNICs from NVIDIA Mellanox and Broadcom, with speeds between 25\,Gbps and 200\,Gbps. Before we build \sys, we already know 3 existing performance anomalies by testing with existing RDMA applications. \sys successfully reproduces all of them and has found 15 new anomalies. We report these anomalies to the vendors, and all of them are acknowledged. 7 of them are already fixed by firmware upgrade or detailed configuration following our vendors' instructions. We also describe our experience in using \sys to guide an RDMA RPC library and an RDMA distributed machine learning framework to avoid these anomalies. These experiences show \sys can help data center operators to uncover anomalies and assist RDMA application developers to implement better applications.

\begin{figure*}[htp]
\centering
\includegraphics[width=0.95\textwidth]{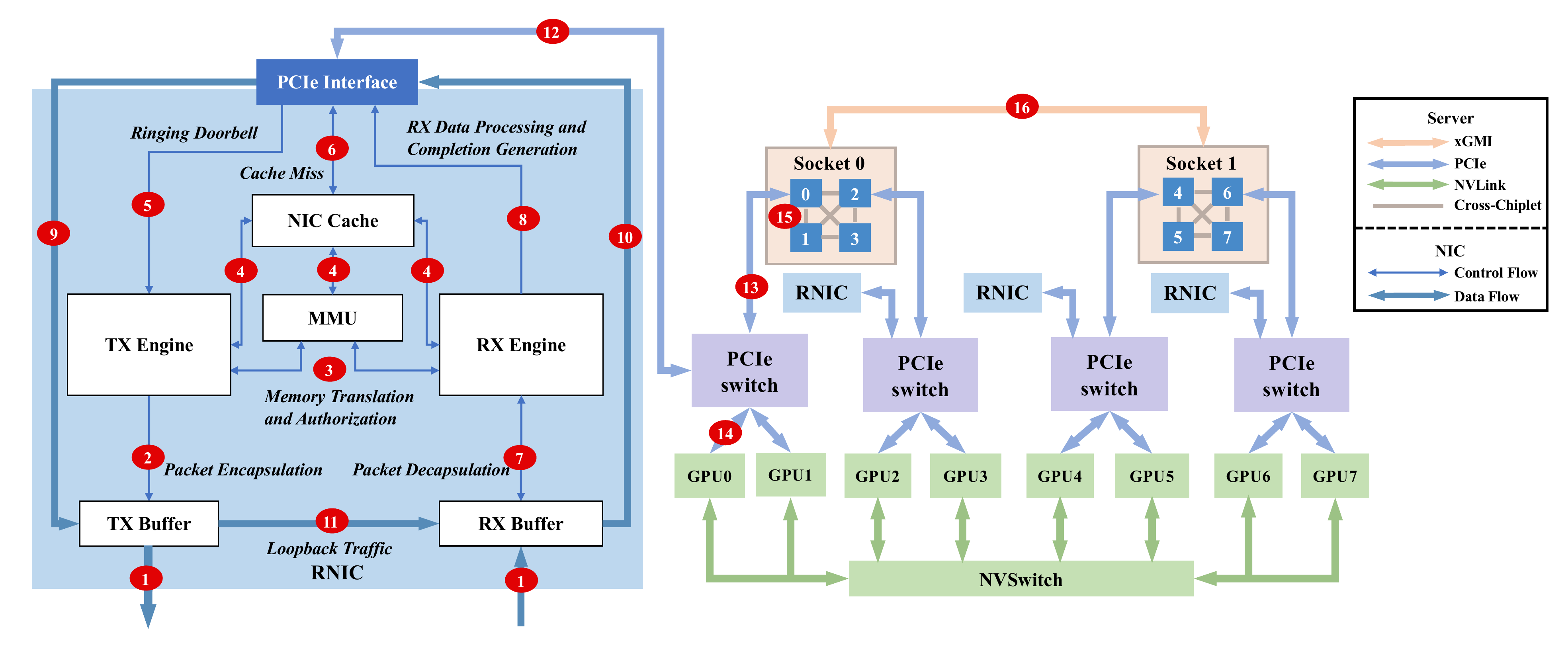}
\vspace{-5mm}
\caption{
An example of an RDMA subsystem (RNIC internal design and its deployment environment in a server). Red circles mean potential performance bottlenecks that can trigger performance anomalies.}
\vspace{-3mm}
\label{fig:server_nic}
\end{figure*}

This work makes the following contributions:
\begin{itemize}
    \item We design a developer-oriented approach to systematically construct a search space of application workloads to find performance anomalies in RDMA subsystems. 
    \item We propose the first work to leverage hardware counters to guide the search for performance anomalies. These counters do not have proprietary hardware knowledge. This makes \sys general and useful for all types of RDMA subsystems. 
    \item We develop a simulated annealing based search algorithm and MFS algorithm. These algorithms speed up search and help developers avoid anomalies.
    \item We implement \sys, the first tool to help data center operators to uncover and avoid RDMA subsystem performance anomalies. \sys has found 18 anomalies (3 known ones and 15 new ones). We present these anomalies, their mitigation strategies, and their implications.
\end{itemize}

\section{Background}
\label{sec:bg}
\subsection{RDMA Subsystem Performance Anomalies}

RDMA is increasingly deployed in data centers for applications to achieve high throughput and high CPU efficiency. An application process can directly communicate through an RNIC with a remote process without involving either side's CPUs. RDMA requires a lossless network to achieve high performance. The default technology to deploy RDMA for Ethernet-based data centers is RoCEv2~\cite{guo2016rdma, zhu2015dcqcn}. It relies on Priority-based Flow Control (PFC)~\cite{pfcrfc} mechanism to guarantee a lossless network: once an ingress queue length exceeds a threshold, the switch/NIC sends out a PFC pause frame to the upstream egress queue, asking the egress queue to pause for a duration to avoid ingress queue overflow. 

RDMA subsystem performance does not always meet user expectations and can have severe performance anomaly. According to our production experience, specific application workloads can trigger hardware bottlenecks of a particular type of RDMA subsystem and cause the entire subsystem performance to drop drastically. Applications of the same subsystem will be affected (e.g., throughput drop) and miss the service level agreement. Worse still, an anomalous RDMA subsystem can send out a large amount of PFC pause frames, which pauses the priority queue of the corresponding switch port and may threaten the entire data center network, such as causing head-of-line blocking and PFC deadlocks ~\cite{guo2016rdma, hu2016pfc, timely_sigcomm_2015}

Though the vendors of RNICs and other hardware components (e.g., GPU, motherboard) have conducted extensive tests on their products, we still find many anomalies in our RoCEv2 production environment. The fundamental reason is that RDMA performance is highly related to the entire RDMA subsystem, consisting of both RNIC internals and other hardware components. \autoref{fig:server_nic} shows the complexity of an RDMA subsystem. \textbf{This figure is based on public resources ~\cite{mlnxprm, 1rma2020sigcomm, storm2019systor} and does not expose proprietary information. Our conversation with Mellanox indicates that a real RNIC is much more complex than our figure shows.} To the best of our understanding, an RNIC has at least 6 components: (1) a \textit{TX engine} that receives doorbells (a signal mechanism for the server to notify RNIC to send a request), fetches and processes requests, and initiates transmission; (2) an \textit{MMU} that translates the virtual address to physical address for RDMA memory regions; (3) an SRAM-based \textit{NIC cache} that caches per-connection metadata and memory translation table; (4) a \textit{RX engine} that processes incoming data and generates completion to notify server; (5)(6) \textit{buffers} that hold packets to transmit and received packets. An RNIC is connected to a server via PCIe. The server has two CPU sockets and each CPU socket has four CPU chiplets (Only AMD CPUs and new-generation Intel CPUs have cross-chiplet communication, otherwise all the cores inside a CPU socket share the last-level cache.) RNICs and GPUs are all connected to PCIe switches. 

There are many potential performance bottlenecks inside the RNIC and between the RNIC and other hardware components within the RDMA subsystem. We use red circles to show such potential bottlenecks (in \autoref{fig:server_nic}). When these bottlenecks are triggered, the network performance may drop and the RNIC can even send out pause frames to reduce the amount of traffic going through the RNIC. We find that many anomalies only occur when multiple bottlenecks or the bottlenecks between different components are triggered. For example, when the RNIC receives a packet, it will store the packet in RX buffer, process the packet (circle 7), and finally DMA the content to main memory or GPU memory (circles 10, 12, 13 or circles 10, 12, 14). Normally, the RX buffer won't accumulate much because the PCIe bandwidth is larger than RNIC's line rate (circle 1). However, once there exists loopback traffic (e.g., the client and server are collocated on the same host), the loopback traffic (circle 11) may drain the PCIe bandwidth and cause RX buffer accumulation. It depends on both the RNIC and the PCIe slot. The worst consequence is that the RNIC keeps sending a large amount of PFC pause frame and threatens the entire data center network. Vendors' individual tests are not able to uncover this anomaly because it depends on the combination of circles 1, 11, 12 (even more) from different components. Further, data center operators like us may use highly customized hardware or specific system configurations that are not accessible to vendors. 
This makes it necessary and crucial for us to conduct our own independent tests before deploying RDMA hardware in production, especially for anomalies that can potentially generate pause frame storms.

\subsection{Existing Approaches}

Data center operators' tests are integration tests: instead of testing individual hardware components, these tests focus on the performance of the entire RDMA subsystems. There are two existing approaches. The first approach is to run a set of test traffic, such as \texttt{Perftest} ~\cite{perftest} and \texttt{OSU micro-benchmarks}~\cite{osubench}. The second approach is to run a representative set of real applications. However, these two approaches can not uncover RDMA subsystem performance anomalies comprehensively. For example, we deploy 200\,Gbps RNICs in our clusters to support a performance-critical distributed machine learning framework. We test the machine learning framework on the cluster of these RNICs, and there is no performance anomaly found. We also have done extensive testing both with synthetic testing workloads and other real applications before deployment. However, months after deployment, our developers find that the performance of the framework has reduced significantly, even worse than just using 100\,Gbps RNICs. At the same time, a substantial amount of pause frames are generated from these 200\,Gbps RNICs. This is strange because pause frames usually appear with hundreds of connections that trigger congestion, but our machine learning framework only creates a few connections between each server pair. We stopped the machine learning framework and ran our performance tests again, and everything is normal. After several weeks of careful debugging, we finally realize that the case only happens when the application (1) use one-sided RDMA operations with Reliable Connection, (2) has bidirectional traffic, (3) uses a particular workload including a mixture of small and large messages, (4) with 200\,Gbps RNIC on particular AMD servers. We find that the developers for our machine learning framework slightly modified their code after passing our application tests. The new workload contains messages of mixed lengths (i.e., a large message followed by a small message followed by a large message in bidirectional traffic), which triggers a performance bottleneck between the RNIC and the PCIe switch. This is not a problem with 100\,Gbps RNICs from the same vendor or on other types of servers. 

\textbf{The fundamental reason} why current approaches fail to uncover such anomalies is that they only test existing workloads and inherently are not able to capture anomalies triggered by unknown workloads. However, real application workloads are various and will change over time. Besides, even current approaches have found such anomalies, it is hard and time-consuming to locate the triggering conditions. Capturing the triggering conditions of performance anomalies allows data center operators to work with vendors to fix potential hardware/firmware bugs, and improve the reliability and performance of the data center network. When fixes to the anomalies are not immediately available (e.g., firmware upgrade, hardware replacement), application developers can build high-performance RDMA applications by avoiding workload that can trigger anomalies. 

\section{Overview}

\begin{figure}
\centering
\includegraphics[width=0.46\textwidth]{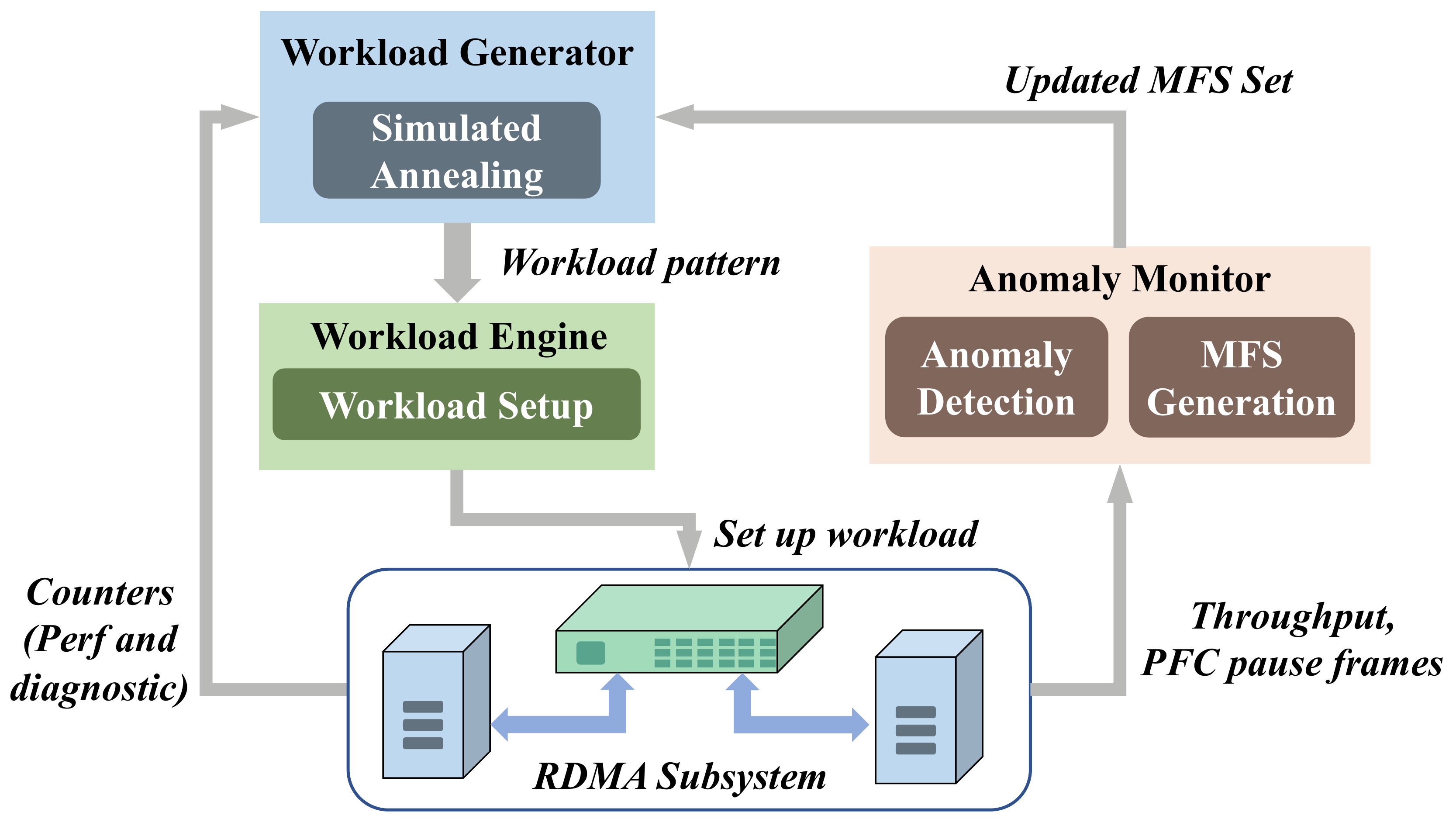}
\vspace{-3mm}
\caption{System Overview. The workload engine sets up RDMA traffic. The anomaly monitor detects performance anomalies and their minimal feature sets.  The workload generator fetches hardware counters and decides the workload pattern to test. 
}
\label{fig:overview}
\end{figure}

We build \sys, the first tool to help data center operators systematically search for application workloads that can trigger performance anomalies.

The first question we need to answer is \textit{how to define an anomaly?} Unfortunately, today there does not exist such a definition. Having such a definition is fundamentally hard because application performance (e.g., latency) can be highly dependent on the workload and the hardware. Instead of trying to capture the entire set of anomalous behaviors, we focus on two types of performance anomalies that are of great importance in production environment and can be precisely defined: when applications keep injecting RDMA traffic into the network, (1) no PFC pause frames should be generated if the network is not congested; (2) throughput should be bottlenecked either by total bits/second or total packets/second as in RNIC specifications. The first definition ensures that an RDMA subsystem will not threaten the entire data center network and the second ensures that an RDMA subsystem's capability matches user expectation. \footnote{We do not use latency as a metric to define anomalies. The only latency specification for RNICs is the latency under zero load. We did not observe any anomaly in this way, probably because the RNIC is not stressed. However, when RNIC is stressed, it is hard to accurately define the correctness of latency or tail latency due to queuing delay.}
We discussed this definition with several hardware vendors, and they all agree with our definition. Even though some anomalies may be due to system limitations rather than bugs, it is also important for both vendors and us to be aware of them. We report all the anomalies we found using this definition to the hardware vendors, and they acknowledged all the reported anomalies. We believe that this definition naturally matches the application developer's mental model of RDMA and thus allows developers to roughly estimate the network performance.

Given this definition of anomaly, we still need to overcome three major research challenges.

\textbf{Challenge \#1:} How to design a comprehensive workloads search space for a given RDMA subsystem? An ideal solution is to carefully analyze and modeling the entire RDMA subsystem, and then construct the search space from the perspective of hardware. This complete white-box approach allows us to test all bottlenecks and the combinations of them givens an RDMA subsystem. However, it is impractical for data center operators like us due to the black-box nature of RNICs and other hardware components. Our key observation is that though the components of RDMA subsystems are black boxes and there are diverse RDMA applications, the abstractions between the hardware and applications are clearly defined and stable. 
All application workloads are essentially composed of a series of basic \textit{verbs} operations, a \textit{narrow waist} of the RDMA programming. With this observation, we carefully analyze this RDMA programming abstraction and design a general search space (\autoref{sec:searchspace}).

\textbf{Challenge \#2:} How to search efficiently? Due to the complexity of RDMA subsystems and the variety of workloads, the size of search space is very large. Unfortunately, none of existing heuristic search algorithms can be directly applied due to the lack of a search signal (e.g., direction for the next workload to test). We observe that there are two sets of counters commodity RDMA subsystem provide can be leveraged to guide the search. The first set is known as performance counters. For example, all modern RNIC provide the counter of bits sent per second for monitoring purpose. The second is known as diagnostic counters. Modern RNICs and other hardware components expose diagnostic counters for debugging purpose (e.g., the counter indicates PCIe backpressure and NIC internal cache miss) ~\cite{diag_counters, neohost}. Diagnostic counters are more informative. For example, when some bottlenecks of the RDMA subsystem are triggered, the performance may not drop while the corresponding diagnostic counter has increased. 
However, using diagnostic counters typically requires vendor's assistance, and the number of diagnostic counters customers can access depends on vendors. For \sys to be general, we use both performance counters and optionally diagnostic counters as search signals. We conduct the efficient search by using a simulated annealing based algorithm to drive these counters to extreme value regions (\autoref{sec:pattern}).

\textbf{Challenge \#3:} How to find the set of conditions to trigger anomalies? Some anomalies are complicated and only occur when many features co-exist, such as a certain type of transportation, particular message pattern, lots of connections, and specific batching operations. We invent a minimal feature set (MFS) algorithm to detect each factor's contribution to the uncovered anomaly and construct the necessary conditions set. To search efficiently, we use MFS to avoid testing similar workloads that map to the same anomalies.
After the search, developers use the MFS to understand the triggering conditions for each anomaly and bypass them accordingly when the fixes are temporarily unavailable (\autoref{sec:anomaly}).

\autoref{fig:overview} shows our system design. \sys consists of three core components: (1) a workload engine that conducts experiments on RDMA subsystems by setting up RDMA traffic; (2) an anomaly monitor that detects performance anomaly and MFS to reproduce the observed anomaly; and (3) a workload generator that decides the next workload pattern to experiment based on the counters collected in the RDMA subsystem and the current search space. All the experiments \sys does are on the RDMA subsystem with two servers with RNICs, connected with a commodity switch.

\section{Search Space and Workload Engine}
\label{sec:searchspace}

There are two types of factors that can affect an RDMA subsystem performance in deployment. First, we need to consider the application workloads. These include host topology (i.e., where does traffic come from inside a server), how many memory regions the application registered, what transport applications choose to use, and the message patterns. Second, we need to consider the network behavior, for example, congestion on switch and packet loss rate. Currently, our paper focuses on constructing a comprehensive search space for the first factor. For the network behavior, we consider a simplified environment: two RNICs connected by a single switch, and there is no packet drop on the switch. \sys can be easily generalized to test more complicated environments.

\begin{figure}
\centering
\includegraphics[width=0.43\textwidth]{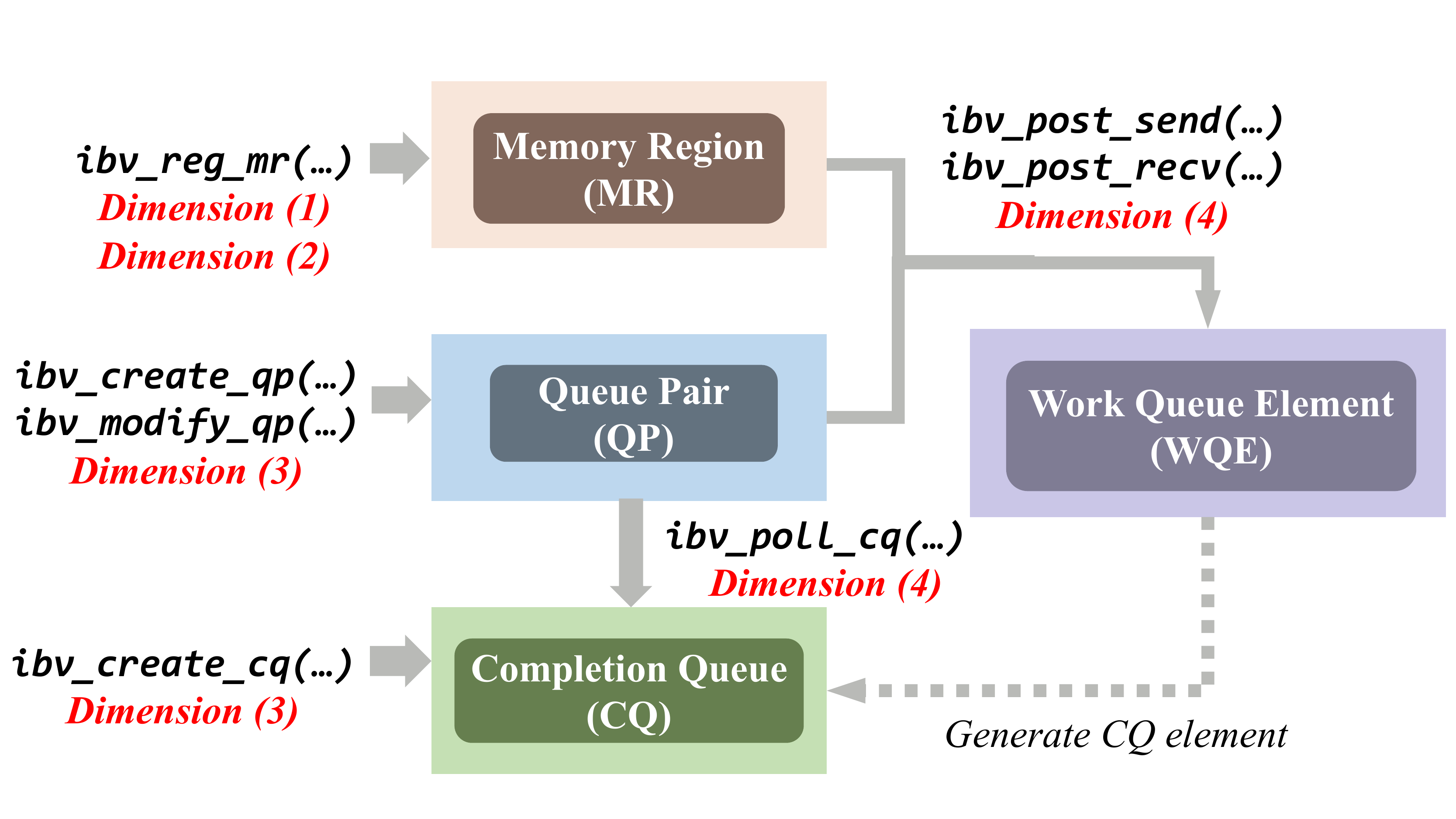}
\vspace{-4mm}
\caption{RDMA programming abstractions}
\label{fig:traffic}
\end{figure}

We take the bottom-up approach to construct a comprehensive search space for various application workloads. We decompose application workloads into combinations of basic RDMA operations and construct the search space based on these combinations. \autoref{fig:traffic} shows the key abstractions and operations of RDMA programming. These are only high-level software abstractions exposed by standard \textit{verbs} API and we do not need to know how these high-level abstractions are implemented in the RNIC. In this way, the search space is more comprehensive and general because it does not rely on either extra proprietary RDMA subsystem hardware knowledge or specific application features. In addition, the combinations of \textit{verbs} operations are inherently able to describe workloads of both single application and co-existing scenarios. 

We examine the RDMA programming model at first and extract four search dimensions that jointly describe the application workloads of the entire subsystem. 
To send a message through RDMA networks, applications first need to register a set of memory regions (MRs), using \texttt{ibv\_reg\_mr}. Once registered, an RNIC has the right to directly access these MRs without CPU involvement. Second, applications create a set of queue pairs (i.e., connections in traditional networking terminology), using \texttt{ibv\_create\_qp} and \texttt{ibv\_modify\_qp}. Applications need to choose a transport type for each queue-pair (QP). There are three standard types of QPs: Reliable Connection (RC), Unreliable Connection (UC), and Unreliable Datagram (UD). Applications can use \texttt{ibv\_post\_send} or \texttt{ibv\_post\_recv} to post a list of Work Queue Element (WQE). Each WQE represents a work request and has a scatter-gather (SG) list. Each SG list contains a list of entries and each entry designates a contiguous memory region that is within the registered memory regions. A WQE can notify the RNIC to READ/WRITE remote memory (1-sided operation) or SEND/RECV local memory to/from a remote server (2-sided operation). To know that a WQE is complete, the application can register a completion queue (CQ) using \texttt{ibv\_create\_cq}, and the application can call \texttt{ibv\_poll\_cq} to poll on the CQ to get completion queue elements (CQE). Given this RDMA programming model, we extract the following search dimensions.

\textbf{Dimension 1. Host Topology.} Host topology describes how traffic flows to/from an RNIC to/from other server hardware components. Individual component tests are hard to cover this dimension while the topology has a huge impact on RDMA subsystem performance. For example, traffic can be from NUMA-affinitive DRAM or from a GPU that needs to traverse both PCIe and SMP interconnect between NUMA nodes. The latter will have a longer data path and therefore higher average DMA latency. This will trigger PCIe backpressure to the RNIC and may induce performance anomalies under some specific application workloads. We list all accessible memory devices for this dimension.

\textbf{Dimension 2. Memory Allocation Settings.} 
Traditional RDMA testing is not comprehensive for this dimension, while the memory allocation settings are crucial for RDMA subsystem performance testing. First, the number of MRs affects RDMA subsystem performance because RNIC has an MMU that translates virtual addresses of memory regions to DMA-capable physical addresses and handles memory \mbox{protection} (e.g., authorization). RNIC only caches a fixed size of entries of the memory address translation table. If too many MRs are registered, it is then likely that the RNIC encounters cache miss and needs to access memory address translation tables on server DRAM via extra PCIe operations. These interactions have an impact on the performance. Second, MRs can have different sizes. This also affects RDMA performance because the size also affects the number of translation table entries. Moreover, many RNICs use Intel Data Direct IO (DDIO) to directly access the CPU's last-level cache. If the access range of an MR is large, it can cause severe cache misses in the CPU's last-level cache~\cite{chen2019scalerpc}.
This dimension is bounded because we can set a reasonable upper bound on the number of MRs (200K), and the MR size is bounded by the total amount of memory that can be registered (pinned) in the physical server.

\textbf{Dimension 3. Transport Setting.} Transportation setting is crucial for RNIC performance, and this is well known in the research community \cite{kalia2016guideline, usingRead2013atc, kalia2019erpc}. 
We use the following factors to compose the transport setting: (1) QP type (RC, UC, UD), (2) the number of QPs, (3) the opcode type (SEND/RECV, WRITE, READ), and (4) the usage of SG and WQE. Different QP type with different opcode creates different pressure for the RNIC. For example, UD does not require ACK for each packet, which lessens the RNIC packet processing pressure. However, the SEND/RECV requires pre-posted receive buffers, which puts more pressure on the RNIC cache. The number of QPs also has a great impact on RNIC performance because of the limited RNIC cache. This is known as the scalability problem~\cite{kalia2019erpc, storm2019systor, chen2019scalerpc}. How SG list and WQE affect RNIC performance is a bit tricky. RNICs have to consume extra PCIe bandwidth to fetch WQE from the host DRAM~\cite{kalia2016guideline}. The PCIe bandwidth consumed by WQE becomes substantial under some particular application workloads and can even be the performance bottleneck. We enumerate all the transport types and the opcodes (e.g., RC WRITE, UD SEND).
It is practical and reasonable to set an upper bound (e.g., 20K) for the number of QPs because data center operators will hardly set up more connections. The SG list and WQE can be parameterized by this formula: $\sum_{i=1}^{n}{m_i} = k$, where $k$ denotes the number of messages to send, $n$ denotes the number of WQE and $m_i$ denotes the number of SG elements within the $i^{th}$ WQE. 

\textbf{Dimension 4. Message Pattern.} Existing RDMA testing approaches lack flexibility and comprehensiveness, especially for this dimension. \texttt{Perftest}~\cite{perftest} only repeatedly send messages of a fixed size and other collective communication benchmarks (e.g., \texttt{OSU benchmark}~\cite{osubench}) test RDMA similarly. These simple benchmark traffic are inadequate for RDMA subsystem testing because they ignore the interaction among different requests (i.e., WQE) in a sequence. 

Our ideal goal is to construct this dimension that can represent any application message pattern. However, it is impractical because application traffic can be very different and the interaction among different requests is unknown given the black-box nature of RNIC. We therefore construct this dimension in the following way. We build a request vector with $n$ elements, where each element describe the request attribute (e.g., size of the message to send). We assume that the $1^{st}$ request affects the $2^{nd}$, the $3^{rd}$, ..., the $n^{th}$ requests but won't affect the request after the $n^{th}$. The larger $n$ we set the larger search space we can cover, but we also need to consume more time. This kind of trade-off is similar to the approach when testing file systems~\cite{martinez2017crashmonkey}, where researchers test fixed-length file system operation sequences. Modern RNIC has limited Processing Units (PU) and pipeline stages~\cite{rdmaturing2021}, restricting the number of outstanding requests an RNIC can process. We thus set $n$ to be the product of the number of PUs and the pipeline stages. We discretize request size into multiple discrete value regions based on MTU and the burst size of the RNIC. The RNIC splits a long request into multiple bursts and processes each burst at one time to avoid Head-of-Line (HoL) blocking. The granularity can be easily modified. With more search time, we can discretize request size in a more fine-grained way. Message inter-arrival time is usually considered as a parameter for application workloads. However, adding the inter-arrival time will substantially extend our search space, so we temporarily only consider the pattern without such inter-arrival time.

\textbf{Workload engine.} We build a flexible workload engine to conduct tests in our search space. Compared to traditional traffic generation tools, e.g., \texttt{Perftest}\footnote{Existing tools, e.g., \texttt{Perftest}, are arguably not designed for this type of testing. They are performance benchmark tools. However, we are not aware of any other tools can that test RDMA subsystems.}, our workload engine is more flexible and has a holistic view. It can send and receive traffic with particular pre-defined patterns (e.g., a large WRITE request followed by a small SEND request). Besides, it supports various memory and transport settings, which can test the entire subsystem holistically. To test with a point in our search space, \sys first translates a test point's settings into a set of input parameters of the workload engine. For example, the setting of dimension 1 and 2 are translated into memory allocation parameters (i.e., which GPU or NUMA DRAM to use and how many MR to register) of the engine. Then, the workload engine will take these input parameters to set up connections and generate traffic.

\section{Search for Performance Anomalies}
\label{sec:search}

The total size of our search space (i.e., the combination of parameters) is on the order of $10^{36}$. Each experiment we do requires 20-60 seconds, mostly depending on the number of QPs to create and the number of MRs to register. This means we cannot exhaust the search space. One naive approach is to generate random input in the search space. This approach is already much better than existing tests because the design of our search space is more comprehensive than that in existing tools (\autoref{sec:eval}). However, similar to typical black-box fuzz testing on software, random inputs can only find few anomalies and cannot efficiently uncover complicated anomalies that require multiple conditions to hold simultaneously.

\subsection{Workloads Generation}
\label{sec:pattern}
We leverage two types of counters to guide the search. The high-level approach is to use an optimization algorithm to drive counters to extreme value regions by keeping mutating the test workloads. For performance counters, we drive the counters to low-value regions. For diagnostics counters (which map to unexpected events), we drive the counters to high-value regions.

\begin{algorithm}[t]
\small
  \caption{Search for Performance Anomalies}
  \begin{algorithmic}[1]
    \Require
      $S$: initial anomaly set;
      $T_0$: a high enough initial temperature;
      $T_{min}$: the lowest limit of temperature;
      $n$: the number of times SA runs for a certain temperature;
    \Ensure
      $S$: An updated anomaly set;
    \State $P_{old}, M_{old} = MeasureRandomPoint()$; pick a random point, setup traffic and collect metrics as $M$
    \While {$T > T_{min}$}
        \For{$i = 0$; $i<n$; $i++$ }
            \State mutate $P_{old} $ for a new application workload $P_{new}$;
            \State \algorithmicif\ {$MatchMFS(S, P_{new}$)} \algorithmicthen\ continue;
            \State $M_{new} = MeasurePoint(P_{new})$;
            \State $\Delta E$ = $CompareMetric(M_{new}, M_{old})$;
            \If {$\Delta$$E<0$} 
                \State $P_{old} = P_{new}$
            \Else 
                \State the probability $prob = exp(-\Delta E/T_{(i)})$;
                \State \algorithmicif\ {$rand(0,1) < prob$ } \algorithmicthen\ $P_{old} = P_{new}$;
            \EndIf
            \If {$IsAnomaly(M_{new})$} 
                \State $new\_mfs = ConstructMFS(P_{new})$;
                \State Put $new\_mfs$ into $S$;
                \State $P_{old}, M_{old} = MeasureRandomPoint()$; pick another random point when a new anomaly is found
            \EndIf
        \EndFor
        \State $T = T * \alpha$; where $\alpha$ is decay factor
    \EndWhile
    \State return $S$ 
  \end{algorithmic}
  \label{alg:sa}
\end{algorithm}
Our algorithm is based on simulated annealing (SA). SA is a probabilistic algorithm to find the global minimum of a given function. The idea is to keep mutating the input in the direction of minimizing a given function. SA calls the function value energy. To avoid getting stuck at a local minimum, SA maintains a temperature value. At the beginning of the algorithm, the temperature is high and SA allows mutating input in the direction of increasing the energy. As temperature decreases during the search, SA is less likely to move the input in the direction of increasing the energy. Finally, when the temperature is below a certain threshold, every mutation of the input must decrease energy. SA finishes when there is no way to mutate the input to make the energy lower.

\autoref{alg:sa} shows our algorithm that is based on SA. We maintain a list of performance anomalies. Each anomaly is an MFS (e.g., an area in the search space) that contains workloads to reproduce the performance anomaly. The search starts from a random workload in the search space, and our algorithm measures the counter values. In each iteration of SA, we mutate the workload in one of our search dimensions (line 4). We test whether the new workload causes a performance anomaly with our anomaly monitor. If so, we run our MFS algorithm to determine the entire area in the search space that belongs to this anomaly. We add the new anomaly to the set and change the current workload to a random one. If the new workload does not trigger a performance anomaly, we measure the point by comparing counter values and decide whether to move the current workload to the new one. We always skip workloads that belong to an existing performance anomaly for efficient search.

Our algorithm extends the standard SA algorithm in several important ways to adapt it for our context. First, we compute the energy in the following way: assuming the counter value changes from A to B, we set the different in energy ($\Delta E$) to be $\frac{B-A}{A}$ for performance counters and $\frac{A-B}{B}$ for diagnostic counters because we are minimizing performance counters and maximizing diagnostic counters to trigger potential anomalies. This also allows us to avoid value region problem (e.g., the value regions of diagnostic counters are sometimes opaque). Second, we do not require SA algorithm to find the actual global optimum because we care about all potential anomalies. We therefore always set a more relaxed temperature and $\alpha$ that enable the algorithm to jump out of a certain stage even when it has already run lots of iterations. In addition, we maintain a set of performance anomalies (i.e., MFS). When mutating the point, we compare the mutated point with our existing MFS (line 5). Each MFS contains a list of parameters ranges. If the mutated point matches all parameters ranges of an MFS (i.e., the parameter value of this point is in the MFS's range), we claim this point belongs to the MFS and skip testing it. This ensures that the future search does not redundantly test workload already covered by the existing set of anomalies.

\begin{table*}[ht]
\small
\centering
\begin{tabular}{|c|c|c|c|c|c|c|c|c|c|}
\hline
    Type & RNIC & Speed & CPU & PCIe & NPS & Memory & GPU & BIOS & Kernel \\
    \hline
    A & CX-5 DX & 25\,Gbps &  Intel(R) Xeon(R) CPU 1 & 3.0 x 16 & 1  & 128 GB & - & INSYDE & 4.19\\
    \hline
    B & CX-5 DX & 100\,Gbps & Intel(R) Xeon(R) CPU 2& 3.0 x 16 & 1 & 768 GB & - & AMI & 4.14\\
    \hline
    C & CX-5 DX & 100\,Gbps & Intel(R) Xeon(R) CPU 2 & 3.0 x 16 & 1 & 384 GB & V100 & AMI & 5.4\\
    \hline
    D & CX-6 DX & 100\,Gbps & Intel(R) Xeon(R) CPU 2& 3.0 x 16 & 1 & 768 GB & - & AMI & 4.14\\
    \hline
    E & CX-6 DX & 200\,Gbps & AMD EPYC CPU 1 & 4.0 x 16 & 1 & 2 TB & A100 & AMI & 5.4 \\
    \hline
    F & CX-6 DX & 200\,Gbps & Intel(R) Xeon(R) CPU 3& 4.0 x 16 & 1 & 2 TB & A100 & AMI & 5.4\\
    \hline
    G & CX-6 VPI & 200 \,Gbps & AMD EPYC CPU 1 & 4.0 x 16 & 2 & 2 TB & - & AMI & 5.4\\
    \hline
    H & P2100G & 100 \,Gbps & Intel(R) Xeon(R) CPU 2  & 3.0 x 16 & 1 & 384 GB & - & AMI & 5.4\\
    \hline
\end{tabular}
\vspace{-2mm}
    \caption{Testbed RDMA subsystems configurations. We use numbers in the name of concrete CPU types for confidentiality.}
\vspace{-5mm}
\label{tab:testbed}
\end{table*}

\subsection{Anomaly Monitor}
\label{sec:anomaly}

Our anomaly monitor detects performance anomalies and computes the MFS of them.

\textbf{Anomaly Detection Condition.} 
We use two conditions to detect anomalies. First, if any pause frame is generated. Here we use a metric called \textit{pause duration ratio}. If the pause duration ratio is 1\%, this means for every second, transmission is paused by 10\,ms. We set our threshold to be 0.1\%. The reason is our experiment platform only has two servers and our switch that connects the servers support line rate traffic, so there is no network congestion to begin with. We set the threshold to be above 0, because RNIC may generate a few pause frames when the memory bus or PCIe bus is busy temporarily, especially when connections are just set up. Second, each RNIC has its maximum bits per second and maximum packets per second in its specification that can be easily verified by running simple benchmarks. Without performance anomalies, network traffic should be restricted by either one of these upper bounds. If a workload's throughput (in terms of both metrics) is 20\% lower than the upper bounds, it means that the performance is likely to be restricted by some other bottlenecks of the RDMA subsystem. \sys reports this and runs the MFS algorithm below.

\textbf{Minimal Feature Set (MFS).} After we detect an anomalous workload, we need to know what features of this workload actually trigger the anomaly. For example, if we currently find a new anomaly that has 5 features. It may be the case that 3 features are already sufficient to reproduce this anomaly.
One approach is to use machine learning based algorithms to generate decision trees or deep neural networks to locate the area in the search space for the anomaly. However, machine learning approaches usually require much more training data and thus many more hardware experiments. 

We instead use a heuristic approach. Since we only have 4 search dimensions with few factors, we just do a few tests on each dimension to determine whether a factor belongs to the MFS. For example, if our search algorithm finds a certain workload using UD can cause a performance anomaly. We test whether the same workload with RC and UC can cause performance anomalies. If not, UD belongs to the MFS because it is necessary to reproduce the anomaly. To determine the MFS of a dimension that is continuous (e.g., number of connections), we discretize them manually into a set of value regions and test each of them. Finer-granularity discretization is acceptable because MFS algorithm only runs when uncovering a new anomaly and the number of anomalies is relatively small compared to the entire search space.

We report all the anomalies to RNIC vendors and we can wait for their fixes. Unfortunately, the solutions to these anomalies are case by case. Some anomalies require vendors to spend a substantial amount of time on coming up with solutions and the solutions may not be applicable for data center operators immediately, such as hardware replacement. Hence, developers need to avoid such anomalies instead of waiting for a fix. 
\sys provides MFS to help developers avoid such anomalies by changing application workload to break the conditions in the MFS. 

MFS helps developers to avoid anomalies in two areas. The first one is anomaly prevention. Before an application is implemented, \sys lets developers restrict the search space using their knowledge of their applications to represent all the possible workloads. Then, \sys outputs whether the restricted search space contains performance anomalies. If not, assuming the developers' understanding of their applications is correct, the application won't encounter any performance anomaly found by \sys. The second one is debugging. When an existing application unfortunately encounters anomalies, we can run \sys on the RDMA subsystem and generate all the MFS. Comparing the application with the generated MFS, \sys provides several suggestions that help to break the triggering conditions. We present two real cases to show how MFS helps developers in \autoref{sec:eval_help_app}.

One caveat of our approach is that we are not able to know the root causes of these anomalies given the black-box nature of the RNICs and other hardware components in the RDMA subsystem. This means it may be the case that multiple MFS are actually due to the same anomaly (i.e., the same hardware bug). 
This is acceptable because the goal of MFS is to accelerate the search algorithm by eliminating redundant test cases and help developers understand what features of the workloads can trigger the anomaly. We anyway need to report all the anomalies (i.e., all the MFS) we found to the vendors and that is also the best we can do given the RNIC black-box hardware nature.

\begin{table*}[!hbpt]
\small
\centering
\begin{tabular}{|c|c|c|c|c|c|c|c|c|c|c|}
\hline
 & RNIC & Direc. & Transport & MTU & WQE & SGE & WQ depth & Message Pattern & \# of QPs & Symptom \\
    \hline
    \cellcolor{green!25} \#1 & CX-6 & - & UD SEND & - & $\ge$64 & - & $\ge 256$ & - & - & pause frame \\
    \hline
    \cellcolor{green!25}\#2 & CX-6 & - & UD SEND & - & $\le$8 & - & $\ge$1024 & $\le$1KB & $\ge\approx$16 & low throup.\\ 
    \hline
    \cellcolor{green!25}\#3 & CX-6 & - & RC READ & 1K & - & - & - & $\ge$16KB & - & pause frame\\
    \hline
    \cellcolor{green!25}\#4 & CX-6 & Bi- & RC READ & - & $\ge$32 & $\ge$4 & - & - & $\ge\approx$160 & pause frame\\
    \hline
    \cellcolor{green!25}\#5 & CX-6 & - & RC SEND & 1K & $\ge$64 & - & $\ge$1024 & $\ge$2KB and $\le$8KB& - & pause frame\\
    \hline
    \cellcolor{green!25}\#6 & CX-6 & - & RC SEND & 1K & $\le$16 & $\ge$2 & $\ge$1024 & $\le$1KB & $\ge\approx$32 & low throup. \\
    \hline
    \cellcolor{green!25}\#7 & CX-6 & - & RC WRITE & - & No & - & - & $\le$1KB and $\ge\approx$12K MRs & - & low throup. \\
    \hline
    \cellcolor{green!25}\#8 & CX-6 & - & RC WRITE & - & No & - & $\le$16 & $\le$1KB & $\ge\approx$500 & low throup. \\
    \hline
    \cellcolor{orange!50}\#9 & CX-6 & Bi- & - & - & - & $\ge $3 & - & mix of $\le$1KB \& $\ge$64KB &-  & pause frame\\
    \hline
    \cellcolor{green!25}\#10 & CX-6 & Bi- & RC WRITE & - & $\ge$64 & - & - & mix of $\le$1KB \& $\ge$64KB & $\ge\approx$320 & pause frame\\
    \hline
    \cellcolor{green!25} \#11 & CX-6 & \multicolumn{8}{c|}{Bidirectional cross-socket traffic on particular AMD servers} & pause frame\\
    \hline
    \cellcolor{orange!50}\#12 & CX-6 & \multicolumn{8}{c|}{Particular GPU-Direct RDMA traffic on particular servers} & pause frame \\
    \hline
    \cellcolor{orange!50}\#13 & CX-6 & \multicolumn{8}{c|}{Co-existence of loop traffic and receiving traffic} & pause frame \\
    \hline
    \cellcolor{green!25} \#14 & P2100 & Bi- & RC & 4K & - & $\ge$4 & - & - & $\ge \approx$1300  & low throup. \\
    \hline
    \cellcolor{green!25} \#15 & P2100 & - & UD SEND & - & - & - & $\ge$64 & - & $\ge \approx$32  & pause frame\\
    \hline
    \cellcolor{green!25} \#16 & P2100 & - & RC READ & 1K & $\ge$8 & - & - & - & $\ge \approx$500  & pause frame\\
    \hline
    \cellcolor{green!25} \#17 & P2100 & - & RC SEND & - & $\le$16 & - & $\ge$128 & $\le$1KB & $\ge \approx$64  & pause frame\\
    \hline
    \cellcolor{green!25} \#18 & P2100 & Bi- & RC & 1K & $\ge$32 & - & - & $\le$64KB & $\ge \approx$30  & pause frame\\
    \hline
\end{tabular}
\vspace{-2mm}
\caption{Performance anomalies found on subsystem F and H with the necessary conditions to trigger them. Anomalies marked with green color are new anomalies found by \sys. Rest are the anomalies we know before building \sys.}
\vspace{-6mm}
\label{tab:anomalies}
\end{table*}

\section{Implementation}

The workload generator and the anomaly monitor are written in \textasciitilde2100 lines of Python. The workload engine is implemented with \textasciitilde2000 lines of C/C++. We directly use monitor tools from vendors to collect hardware counters (both performance and diagnostic counters) from the RDMA subsystem.

The workload engine is implemented with the \textit{verbs} API and \textit{rdma-core-34.0} libraries\cite{rdma-core}. In deployment, the Mellanox RNIC uses mlx5 driver (\CR{OFED 5.2-1.0.4.0}) and the Broadcom RNIC uses bnxt driver (1.10.1.216.2.89.0). The workload engine set up connections by TCP out-of-band transmission. When all connections are set up, the engine starts to generate workload.

The anomaly monitor collects primary metrics, such as throughput and pause frame duration, four times per iteration. It first decides whether the traffic is stable and then compares the primary metrics (e.g., bits per second, packets per second) with the pre-defined thresholds. 
 
The workload generator collects counters using monitors provided by vendors. These monitors provide counters every second. \sys fetches these counters four times per iteration and uses the average results. 

\section{Evaluation and Experience}
\label{sec:eval}

We evaluate \sys on 8 different RDMA subsystems. \autoref{tab:testbed} shows the hardware and related configurations. We use the same anomaly detect conditions as described in \autoref{sec:anomaly}

\subsection{Performance Anomalies Found}
Before we build \sys, we already know 3 existing anomalies. \sys can find all the existing ones and find 15 new anomalies. All of them are reported to our vendors and are acknowledged by them. \autoref{tab:anomalies} shows the 18 anomalies. We only present those found on subsystem F and H because anomalies found on other subsystems are subsets of those found on F. \autoref{app:anomalies} provides details about these anomalies, including the exact workload, \CR{as well as} the explanations and solutions from vendors. Here we choose two tricky anomalies to show the importance of \sys's systematic search.

\textbf{Anomaly \#4:} Bidirectional RC READ with large WQE batch size, long SG list, and a few connections causes PFC pause frames. Our vendors have successfully reproduced this anomaly in their environment \CR{using \sys's traffic generator} and acknowledged it, but currently there is no fix. This anomaly cannot be found by existing approaches such as using \texttt{Perftest} to generate workloads, because \texttt{Perftest} does not support flexible WQE and SG list batching strategies. Though \texttt{Perftest} is not designed for this purpose, it is the prevalent tool to uncover performance anomalies. To the best of our knowledge, we don't see any other state-of-the-art work address this problem, which also shows that \sys is the first work to fill this vacancy.

\textbf{Anomaly \#10:} Bidirectional RC WRITE with large WQE batch size, particular message pattern, and a few connections causes PFC pause frames. This anomaly is not captured by existing approaches (e.g., running current applications) but we successfully reproduce it by slightly modifying our production RDMA RPC library: when users call the library to send a message, it will try to send as many messages as possible in a WQE batch. The batch size is highly dependent on the timeout value. If the application is throughput sensitive rather than latency sensitive, the timeout value can be set high, which allows a larger batch size. Currently the timeout value is set small because most applications supported by this library are latency sensitive. However, by changing this value we successfully enlarge the WQE batch size and the conditions of \#10 are all met. This shows the importance of the anomalies found by \sys, as well as how \sys can capture those anomalies missed by existing solutions.  

We try our best to reproduce the anomalies found by \sys using existing workload generators (e.g., \texttt{Perftest}), only 4 of them (\#3, \#8, \#13, \#15) can be reproduced with very careful parameters tuning. Rest anomalies are all outside the search space of existing approaches.

\subsection{Running Time for Anomaly Search}

To evaluate the efficiency of performance anomaly search, we compare \sys with two baselines: (1) random input generation in our search space and (2) Bayesian Optimization (BO), a widely used method in search problem ~\cite{BO_github}. We implement the BO approach based on ~\cite{BO_github}. \CR{We set the counter values as BO's optimization target}. Our vendors provide us with 9 diagnostic counters. For \sys and BO, we first generate 10 random points. We then compute the standard deviation over the mean of the counter values collected in the first 10 run and use the result to rank these diagnostic counters in decreasing order. Both \sys and BO optimize each diagnostic counter in this order. For a fair comparison, we use MFS to enhance BO as well. In this section, we use subsystem F as an example. We run each search for 10 hours.

\begin{figure}[t]
\centering
\includegraphics[width=0.46\textwidth]{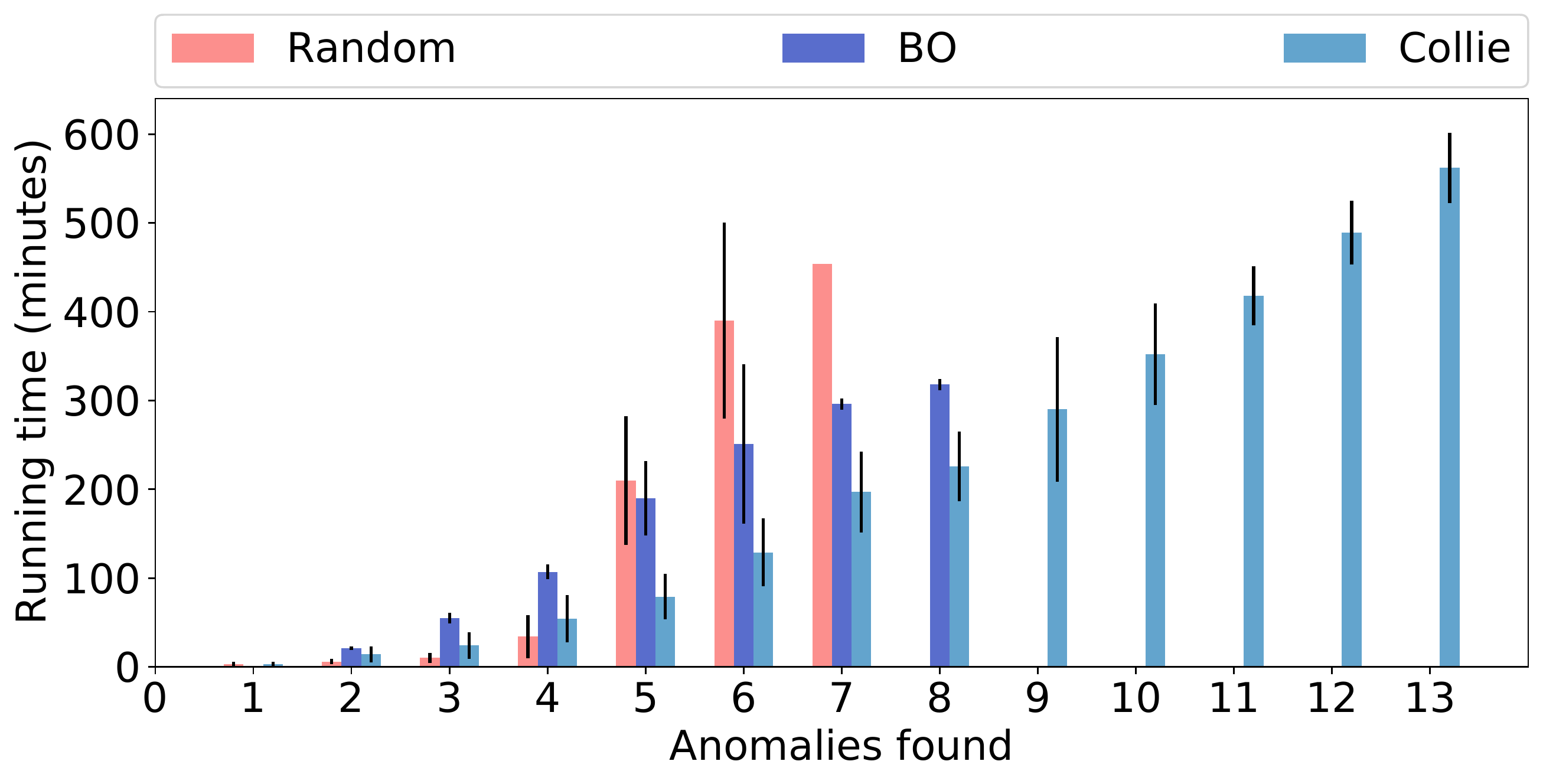}
\vspace{-4mm}
\caption{Mean time to find anomalies with random input generation, BO, and \sys. Error bars denote standard deviations. There is no red bar starting from 8, and no purple bar starting from 9, because random input generation and BO can only find 7 and 8 anomalies, respectively.}
\label{fig:anomaly_count}
\end{figure}

\CR{\autoref{fig:anomaly_count} shows the running time to find performance anomalies. Random input (i.e., fuzzing) can already find 7 anomalies that only require simple conditions to trigger. BO does improve efficiency but to a very limited extent. BO can speed up the search process but only find 8 anomalies with the given time. We analyze the optimization process of BO and find that it is not able to optimize the corresponding counters. Our guess is that BO works well when counter values are smooth in the search space. However, the counter values in our search space can have sudden changes, because some discrete dimensions have a huge impact on the counter values (e.g., QP type). \sys uses a simulated annealing based algorithm to optimize the counter values and successfully speed up the search process. Given limited time, it can find all the performance anomalies of this RDMA subsystem. We believe this improvement comes from the optimization process: driving counters to extreme regions is more likely to trigger performance anomalies. It is possible that a more efficient search algorithm (e.g., a fine-tuned BO, reinforcement learning) can perform better, and it is worth future exploration. However, our goal here is to demonstrate that existing simple optimization algorithms, such as simulated annealing, can search efficiently with these hardware counters.}

\begin{figure}[t]
\centering
\includegraphics[width=0.46\textwidth]{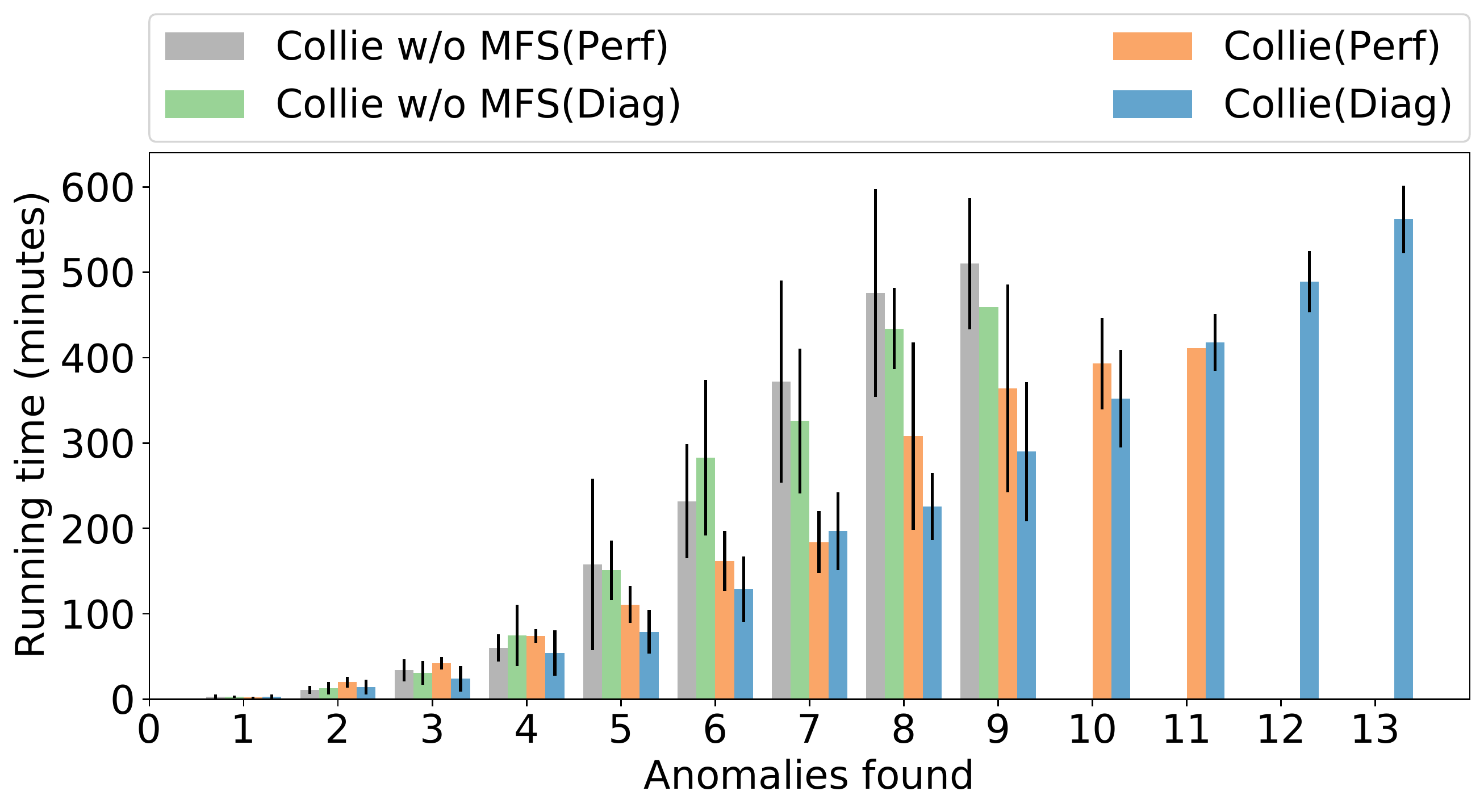}
\vspace{-4mm}
\caption{Mean time to find anomalies. (Diag) means diagnostic counters, and (Perf) means performance counters. Error bars denote standard deviations. }
\label{fig:mfs_perf_diag}
\end{figure}

\sys uses diagnostic counters and MFS to further speed up the search. Now we break down their contribution to our overall search speed. \autoref{fig:mfs_perf_diag} shows the result. 

\textbf{The value of diagnostic counters.} \autoref{fig:mfs_perf_diag} shows that with performance counters, \sys(Perf) has already found 11 of the 13 anomalies, including the 3 existing ones. This proves that the performance counters are informative and can be used to improve search efficiency. It shows the generality of \sys because performance counters are general and provided by all commodity RDMA subsystems. \autoref{fig:mfs_perf_diag} also shows that using diagnostic counters can further improve the speed. Given limited time, \sys(Diag) can uncover more anomalies and is faster. For example, Anomalies \#7 and \#8 are not captured by \sys(Perf) because there is no performance change during the search, but \sys(Diag) can observe the increase of RNIC internal cache miss and uncover them.

\textbf{The value of minimal feature set (MFS)}
The main difference between SA and \sys is whether MFS is applied. With MFS, the efficiency of all approaches (both using diagnostic counters and using performance counters) is significantly improved. For example, \sys(Diag) only uses about half of the time to uncover all the anomalies found by SA(Diag). MFS improves efficiency by eliminating redundant tests from the search space. Otherwise, approaches without MFS may be stuck in the area of an uncovered anomaly.  

To understand why increasing diagnostic counter values can help to find anomalies and how MFS works, here we use \textit{Receive WQE Cache Miss} counter as an example. We do not rely on the meaning of these diagnostic counters during the search. To the best of our knowledge, the counter means the number of times that RNICs need to issue extra DMA operations to fetch receive WQE from host DRAM.

\begin{figure*}[t]
\centering
\includegraphics[width=0.85\textwidth]{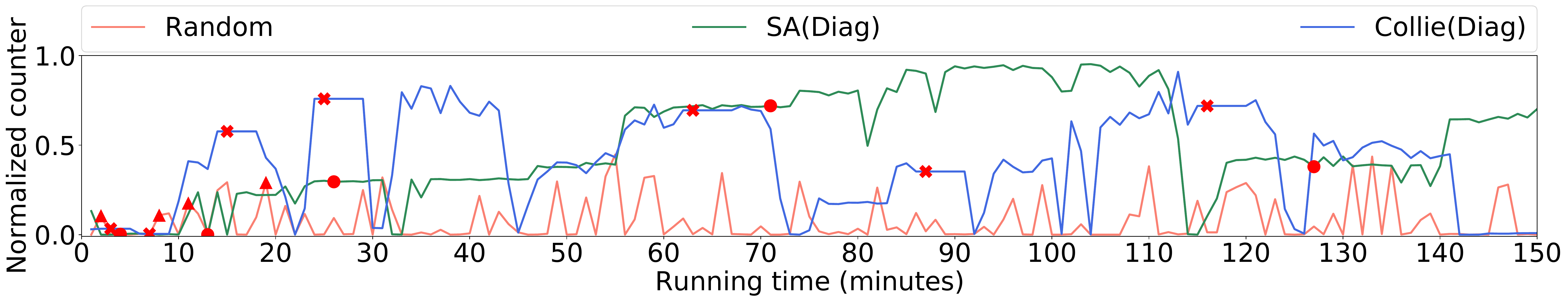}
\vspace{-4mm}
\caption{Diagnostic counter values (\textit{Receive WQE Cache Miss}) during the search. Counter values are normalized based on the maximum value we observed in the search. Red crossings denote the performance anomalies found by \sys. Red triangles denote the performance anomalies found by random input generation. Red squares denote the performance anomalies found by \sys without MFS. \sys (the blue line) is flat for a few minutes after finding a new performance anomaly. This is to represent the time needed for extracting the MFS.}
\vspace{-3mm}
\label{fig:key_graph}
\end{figure*}

\autoref{fig:key_graph} shows the diagnostic counter values during the search. The random input generation approach (the orange line) does not increase the diagnostic counter value and thus cannot find many performance anomalies. \sys w/o MFS (the green line) can drive the diagnostic counter value very high, but it cannot find many distinct performance anomalies because further increasing the counter value in the neighboring regions of existing performance anomalies wastes time. \sys (the blue line) is effective in finding performance anomalies, because it can both increase the diagnostic counter value to find application workloads that cause anomalies and also do not need to test application workloads that belong to the same anomaly. \autoref{fig:key_graph} shows that most anomalies are found when the diagnostic counter value is high. This also supports the intuition that it is likely to trigger performance anomalies when the diagnostic counter value is driven to extreme regions, which indicates the RDMA subsystem is under pressure. Some anomalies in \autoref{fig:key_graph} do not show a high value of this counter. This is mainly due to that they are anomalies that can be easily triggered. They are usually triggered at the beginning of the search process (left corner of \autoref{fig:key_graph}) and another corresponding diagnostic counter value is high. For example, Anomaly \#13 has simple triggering conditions and is usually found very soon. It does not increase the \textit{Receive WQE Cache Miss} counter but will increase another counter, the counter of \textit{PCIe Internal Back Pressure}. 

\subsection{Using \sys for Application Design}
\label{sec:eval_help_app}

We use \sys in the development and performance debugging of two key RDMA applications.

First, \sys provides design suggestions for \CR{our self-developed} efficient RDMA RPC library during its design and implementation. The library needs to be CPU-efficient, and we thus only consider RC as the transport because it is the only transport that supports all one-sided RDMA operations (i.e., READ, WRITE) and ensures reliable messages. \CR{In addition, major services that use this RPC library will mainly be deployed on subsystem B and C.} Given the search space, \sys provides two suggestions to the developers. (1) Anomaly \#4 is in the restricted search space if the RDMA RPC library uses READ, large WQE batch size, and a long SG list to improve throughput and shape the message format. (2) The library needs to use SEND/RECV to deliver small control messages and generally keeps a large receive queue in case of receive-not-ready error. This can potentially trigger Anomaly \#5. Unfortunately, both \#4 and \#5 temporarily have no fix, so \sys suggest developers (1) use RDMA WRITE to transmit data in a batch and (2) configure receive queue depth carefully in SEND/RECV for small control messages transmission. This RDMA based RPC library achieves expected performance and is currently supporting three major services in production.

\CR{Second, \sys helps a distributed machine learning (DML) application based on BytePS~\cite{jiang2020byteps} bypass anomalies during its further development in our production environment. Our DML application encountered anomaly \#9 when deploying on our new subsystem E. We worked with multiple vendors (RNIC, server, CPU), but for several weeks we didn't find the root cause or the fix for this anomaly. During this time, we ran \sys and compared the anomalous application with the MFS we got. We found that the application's behaviors matched one of the MFS: (1) use a long SG list to send tensors with several meta data and (2) the message pattern of tensors and meta data is a typical pattern that contains mix of short and long messages. \sys suggested the developers to avoid these conditions. The developers hence  bypassed this anomaly before vendors' fix is ready. }

\subsection{Implications of the Performance Anomalies Found}
\label{sec:eval_implications}

After careful analysis of the anomalies found by \sys, we have several interesting and important observations.

\textbf{Holistic performance testing/tuning over entire RDMA subsystems is important.} With our vendors' help, we try our best effort to present the root causes of these anomalies in \autoref{app:anomalies}. The root causes can be bottlenecks from RNIC internals, PCIe controllers, and host topologies (cross socket communication). This is because the RDMA network performance is highly related to the entire subsystem and the holistic test is thus important. Besides, we need to configure systems carefully  (MTU, PCIe, NUMA, IOMMU, etc.) to fully leverage RDMA's performance~\cite{neugebauer2018pcie, kalia2016guideline}. \sys shows that it is sometimes difficult to choose what configuration to use. For example, comparing the Anomaly \#14 with other cases related to the MTU setting (e.g., \#6), we observe there is no optimal MTU setting for all types of RDMA subsystems. This also indicates that data center operators have to test various RDMA subsystem configurations and tune the system carefully before deploying them.  

\textbf{Opaque resource limitation of the RDMA subsystems.} RDMA virtualization, especially performance isolation is important for deploying RDMA to the public cloud environment. Researchers have spent a lot of effort and proposed several solutions~\cite{lite2017, freeflow_2019, zhang2019rdma, masq_2020, hyv_2015_vee}. However, anomalies found by \sys suggest that there are new challenges. Existing approaches mainly focus on the isolation of visible resources like \textit{verbs} structures (e.g., QP, MR, CQ), pinned memory, and bandwidth. However, there exist resources that are opaque for developers and data center operators. For example, the RNIC has limited caches that store many data structures, including connection context (well known as QPC) and receive WQE. Anomalies \#1, \#3, \#4, \#5 show that severe WQE cache miss can have a huge impact on performance. Hence, it is possible that a connection with a specific message pattern affects another connection by triggering cache misses, even when the bandwidth and other resources are well isolated. We therefore believe it is necessary to take these invisible resources into consideration when enforcing RDMA performance isolation, especially in public clouds.

\textbf{Does Ethernet-based RDMA need end-to-end flow control?} Currently there is no end-to-end flow control mechanism (e.g., the sliding window for TCP) for production Ethernet-based RDMA deployment (i.e., RoCEv2). \sys shows that this is a major barrier for RDMA subsystems to achieve high-performance and reliability. For example, many anomalies (e.g., \#9 and \#12) show that the host limitation can slow down RNIC's outbound rate (dispatching received data to host memory). This makes the receiver cannot consume packets as fast as the sender sends. Without end-to-end flow control, the RoCEv2 now can only rely on PFC, the hop-by-hop flow control mechanism. PFC helps to avoid such overflow packet drop but can cause catastrophic consequences~\cite{guo2016rdma, hu2016pfc}. Note that RDMA congestion control ~\cite{zhu2015dcqcn, timely_sigcomm_2015, li2019hpcc} mainly targets in-network congestion, so it is orthogonal. A similar observation has been shown in IRN \cite{irn_sigcomm_2018}, but they mainly focus on in-network behaviors.
\sys shows that, in addition to switches, the hosts can also generate PFC pause frames, which requires attention when deploying RDMA in production.
\section{Discussion and Future Work}

\textbf{Search space.}
\sys mainly focuses on how specific application workloads can \CR{stress} the RDMA subsystems and trigger performance anomalies. \CR{We therefore focus on a simple setting of two RNICs and assume the network is free of anomaly. In addition, we temporarily ignore control path behaviors and the inter-arrival time between requests of a connection. The main reason is that adding these factors substantially enlarge the size of our search space.} 
How to efficiently expand \sys's search space is an interesting direction for exploration.

\textbf{Search algorithm.}
\sys uses simulated annealing based algorithm with minimal feature set (MFS) to search efficiently. Though powerful data centers can run \sys on multiple machines for a longer time, the search algorithm is also important. According to the MFS found by \sys, the expected time for a random approach is tens of days to find some anomalies that require complicated triggering conditions. There are many other search algorithms alternatives that can be leveraged, such as reinforcement learning. Integrating more search algorithms into \sys is another interesting direction to explore. 

\textbf{Generality of \sys.}
We believe that \sys can be used for any type of RDMA subsystem or even subsystems with other types of NICs. For example, though the link/transport protocols are different for Infiniband and RoCEv2, the NIC internal structures should be similar (e.g., both can use Mellanox CX-6 VPI RNIC). \sys only relies on non-proprietary counters that expose NIC internal status. Therefore, this methodology should be generalizable to any NIC in any deployment environment if similar counters are available.

\CR{\textbf{Analysis of Performance Anomalies.}
\sys is designed to uncover anomalies and help to bypass them from the perspective of data center operators, so it assumes minimal hardware knowledge of RDMA subsystems for generality and does not directly analyze the underlying causes. However, since the anomalies found by \sys can be severe (e.g., triggering PFC pause storms), we believe to fully understand them is also an important direction to explore. For example, as mentioned in \autoref{sec:eval_implications}, many anomalies are due to bottlenecks on some opaque resources. Both RNIC vendors and data center operators hence need to understand what extra resources should be considered if they want to provide performance isolation for RDMA in a public cloud. 

}
\section{Related Work}
\textbf{Hardware bottlenecks in host networking.} 
With the fast growth in NIC performance, researchers have noticed several potential hardware bottlenecks in host networking. Neugebauer et al.~\cite{neugebauer2018pcie} study the implication of PCIe performance in host networking. Farshin et al.~\cite{farshin2020ddio} examine when and when not Intel Data Direct I/O technology can speed up host networking by allowing NIC to access CPU's last-level cache directly. Kalia et al.~\cite{kalia2019erpc} observe the scalability bottlenecks of caching per-connection metadata in RNIC. Stanko et al.~\cite{storm2019systor} study how the number of connections and memory regions affect performance.
These works have raised our attention to RNIC hardware behaviors. Our work is on a different angle: we systematically uncover the performance anomalies that can be triggered by specific application workload due to hardware bottlenecks.

\textbf{Fuzz testing.} Our techniques are in the broader category of fuzz testing. There are three types of fuzz testing: black-box~\cite{miller1990blackbox, miller2020fuzzing}, white-box~\cite{godefroid2008whitebox, Godefroid2008AutomatedWF, ganesh2009whitebox}, and gray-box fuzzing~\cite{bhome2017graybox, schmilo2017graybox}. Black-box fuzzing is to generate random inputs to test a program, and usually black-box fuzzing can only uncover shallow bugs. In our context, this is also true that using randomly generated application workload can only uncover a small set of anomalies (\autoref{sec:eval}). White-box fuzzing is to use symbolic execution on source code to guide the fuzzer to generate inputs that can have high coverage. We do not have the internal designs of the various components within an RDMA subsystem, so we cannot use white-box approaches. Gray-box fuzzing in the software context is to use the coverage in the control flow graph to guide the fuzzer to incrementally generate inputs that can lead to larger coverage. Our approach is similar to gray-box fuzzing that we both use simulated annealing and mutation-based test case generation. However, the key difference is that we use hardware counters in the RDMA subsystem to guide the search rather than the coverage on the control flow graphs of the source code.

\textbf{Application design on top of RDMA.} Many RDMA application designs leverage specific RDMA performance characteristics, and some already try to circumvent certain RNIC performance anomalies. HERD~\cite{kalia2014herd} uses UD SEND and UC Write to implement an RPC library for reduced RNIC packet processing overheads and better scalability.
FaSST~\cite{kalia2016fasst} and eRPC~\cite{kalia2019erpc} uses UD to further mitigate RNIC scalability bottlenecks in RPC libraries.
Kalia et al.~\cite{kalia2016guideline} provide guidelines to optimize HERD's transport by considering PCIe bottlenecks.
FaRM~\cite{aleksandar2014farm, dragojevic2015farm} uses RC to access remote in-memory key-value stores, so that it can use RDMA 1-sided READ/WRITE operation for reduced CPU overheads. Our goal is complementary: we systematically uncover the set of performance anomalies of RDMA subsystems that application developers need to be aware of. We show that for RDMA developers, in reality, there is no optimal choice for a particular design decision (e.g., all transport types have certain performance anomalies). Developers therefore need to have a holistic view of all the design decisions and the entire RDMA subsystem before designing and implementing RDMA applications.

\section{Conclusion}
RDMA has been increasingly used in the industry for its low latency and reduced CPU overheads. Performance anomalies hurt application performance and can lead to catastrophic consequences (e.g., deadlocking the data center network).
We build \sys, a tool to help RDMA users to find performance anomalies of the entire RDMA subsystems, without the need for access to any hardware internals design.
\sys constructs a comprehensive search space for RDMA application workloads and finds performance anomalies by using simulated annealing to optimize two types of vendor-provided counters. We evaluate \sys on 8 commodity RDMA subsystems and \sys found 15 new performance anomalies that are all acknowledged by the vendor. 7 of them are already fixed under vendors' guidance. We also present our experience in using \sys to guide our development of an RDMA RPC library and help our distributed machine learning applications bypass performance anomalies before vendor fix is ready. Collie is available at \url{https://github.com/bytedance/Collie}.

\section*{Acknowledgement}
We thank Alvin R. Lebeck, Xiaowei Yang, Xi Wang, Wei Bai, Mahmoud Elhaddad, Jitu Padhye, and Shachar Raindel for their helpful comments and discussion. We thank NVIDIA, Broadcom, and AMD for their strong technical support. We thank our shepherd Costin Raiciu and other anonymous reviewers for their insightful feedback. Our work is partially supported by an Amazon Research Award, a Meta Research Award, and an IBM Academic Award. 

\clearpage
\bibliographystyle{plain}
\bibliography{reference}

\begin{thebibliography}{10}

\bibitem{bhome2017graybox}
Marcel B\"{o}hme, Van-Thuan Pham, Manh-Dung Nguyen, and Abhik Roychoudhury.
\newblock {Directed Greybox Fuzzing}.
\newblock In {\em CCS}, 2017.

\bibitem{buragohain2020a1}
Chiranjeeb Buragohain, Knut~Magne Risvik, Paul Brett, Miguel Castro, Wonhee
  Cho, Joshua Cowhig, Nikolas Gloy, Karthik Kalyanaraman, Richendra Khanna,
  John Pao, et~al.
\newblock {A1: A Distributed In-Memory Graph Database}.
\newblock In {\em SIGMOD}, 2020.

\bibitem{chen2019scalerpc}
Youmin Chen, Youyou Lu, and Jiwu Shu.
\newblock {Scalable RDMA RPC on Reliable Connection with Efficient Resource
  Sharing}.
\newblock In {\em EuroSys}, 2019.

\bibitem{aleksandar2014farm}
Aleksandar Dragojevi{\'c}, Dushyanth Narayanan, Miguel Castro, and Orion
  Hodson.
\newblock {FaRM: Fast Remote Memory}.
\newblock In {\em NSDI}, 2014.

\bibitem{dragojevic2015farm}
Aleksandar Dragojevi\'{c}, Dushyanth Narayanan, Edmund~B. Nightingale, Matthew
  Renzelmann, Alex Shamis, Anirudh Badam, and Miguel Castro.
\newblock {No Compromises: Distributed Transactions with Consistency,
  Availability, and Performance}.
\newblock In {\em SOSP}, 2015.

\bibitem{farshin2020ddio}
Alireza Farshin, Amir Roozbeh, Gerald Q.~Maguire Jr., and Dejan Kosti{\'c}.
\newblock {Reexamining Direct Cache Access to Optimize I/O Intensive
  Applications for Multi-hundred-gigabit Networks}.
\newblock In {\em USENIX ATC}, 2020.

\bibitem{ganesh2009whitebox}
Vijay Ganesh, Tim Leek, and Martin Rinard.
\newblock {Taint-Based Directed Whitebox Fuzzing}.
\newblock In {\em ICSE}, 2009.

\bibitem{yixiao2021nsdi}
Yixiao Gao, Qiang Li, Lingbo Tang, Yongqing Xi, Pengcheng Zhang, Wenwen Peng,
  Bo~Li, Yaohui Wu, Shaozong Liu, Lei Yan, Fei Feng, Yan Zhuang, Fan Liu, Pan
  Liu, Xingkui Liu, Zhongjie Wu, Junping Wu, Zheng Cao, Chen Tian, Jinbo Wu,
  Jiaji Zhu, Haiyong Wang, Dennis Cai, and Jiesheng Wu.
\newblock {When Cloud Storage Meets {RDMA}}.
\newblock In {\em NSDI}, 2021.

\bibitem{godefroid2008whitebox}
Patrice Godefroid, Adam Kiezun, and Michael~Y. Levin.
\newblock {Grammar-Based Whitebox Fuzzing}.
\newblock In {\em PLDI}, 2008.

\bibitem{Godefroid2008AutomatedWF}
Patrice Godefroid, Michael~Y. Levin, and D.~Molnar.
\newblock {Automated Whitebox Fuzz Testing}.
\newblock In {\em NDSS}, 2008.

\bibitem{guo2016rdma}
Chuanxiong Guo, Haitao Wu, Zhong Deng, Gaurav Soni, Jianxi Ye, Jitu Padhye, and
  Marina Lipshteyn.
\newblock {RDMA over Commodity Ethernet at Scale}.
\newblock In {\em SIGCOMM}, 2016.

\bibitem{masq_2020}
Zhiqiang He, Dongyang Wang, Binzhang Fu, Kun Tan, Bei Hua, Zhi-Li Zhang, and
  Kai Zheng.
\newblock {MasQ: RDMA for Virtual Private Cloud}.
\newblock In {\em SIGCOMM}, 2020.

\bibitem{hu2016pfc}
Shuihai Hu, Yibo Zhu, Peng Cheng, Chuanxiong Guo, Kun Tan, Jitendra Padhye, and
  Kai Chen.
\newblock {Deadlocks in Datacenter Networks: Why Do They Form, and How to Avoid
  Them}.
\newblock In {\em HotNets}, 2016.

\bibitem{jiang2020byteps}
Yimin Jiang, Yibo Zhu, Chang Lan, Bairen Yi, Yong Cui, and Chuanxiong Guo.
\newblock {A Unified Architecture for Accelerating Distributed {DNN} Training
  in Heterogeneous GPU/CPU Clusters}.
\newblock In {\em OSDI}, 2020.

\bibitem{kalia2019erpc}
Anuj Kalia, Michael Kaminsky, and David Andersen.
\newblock {Datacenter RPCs can be General and Fast}.
\newblock In {\em NSDI}, 2019.

\bibitem{kalia2014herd}
Anuj Kalia, Michael Kaminsky, and David~G. Andersen.
\newblock {Using RDMA Efficiently for Key-Value Services}.
\newblock In {\em SIGCOMM}, 2014.

\bibitem{kalia2016guideline}
Anuj Kalia, Michael Kaminsky, and David~G. Andersen.
\newblock {Design Guidelines for High Performance {RDMA} Systems}.
\newblock In {\em USENIX ATC}, 2016.

\bibitem{kalia2016fasst}
Anuj Kalia, Michael Kaminsky, and David~G. Andersen.
\newblock {FaSST: Fast, Scalable and Simple Distributed Transactions with
  Two-Sided ({RDMA}) Datagram RPCs}.
\newblock In {\em OSDI}, 2016.

\bibitem{freeflow_2019}
Daehyeok Kim, Tianlong Yu, Hongqiang~Harry Liu, Yibo Zhu, Jitu Padhye, Shachar
  Raindel, Chuanxiong Guo, Vyas Sekar, and Srinivasan Seshan.
\newblock {FreeFlow: Software-based Virtual {RDMA} Networking for Containerized
  Clouds}.
\newblock In {\em NSDI}, 2019.

\bibitem{li2019hpcc}
Yuliang Li, Rui Miao, Hongqiang~Harry Liu, Yan Zhuang, Fei Feng, Lingbo Tang,
  Zheng Cao, Ming Zhang, Frank Kelly, Mohammad Alizadeh, and Minlan Yu.
\newblock {HPCC: High Precision Congestion Control}.
\newblock In {\em SIGCOMM}, 2019.

\bibitem{martinez2017crashmonkey}
Ashlie Martinez and Vijay Chidambaram.
\newblock {CrashMonkey: A Framework to Automatically Test File-System Crash
  Consistency}.
\newblock In {\em HotStorage}, 2017.

\bibitem{diag_counters}
Mellanox.
\newblock {Device Proprietary Counters}.
\newblock
  \url{https://docs.nvidia.com/networking/display/WINOFv55052000/Device+Proprietary+Counters}.

\bibitem{neohost}
Mellanox.
\newblock {NEO-Host}.
\newblock
  \url{https://support.mellanox.com/s/productdetails/a2v50000000N2OlAAK/mellanox-neohost}.

\bibitem{mlnxprm}
{Mellanox Adapters Programmer’s Reference Manual (PRM)}.
\newblock
  \url{https://www.mellanox.com/related-docs/user_manuals/Ethernet_Adapters_Programming_Manual.pdf},
  2021.

\bibitem{miller2020fuzzing}
Barton Miller, Mengxiao Zhang, and Elisa Heymann.
\newblock {The Relevance of Classic Fuzz Testing: Have We Solved This One?}
\newblock {\em IEEE Transactions on Software Engineering}, page 1–1, 2020.

\bibitem{miller1990blackbox}
Barton~P. Miller, Louis Fredriksen, and Bryan So.
\newblock {An Empirical Study of the Reliability of UNIX Utilities}.
\newblock {\em Commun. ACM}, 33(12):32–44, December 1990.

\bibitem{usingRead2013atc}
Christopher Mitchell, Yifeng Geng, and Jinyang Li.
\newblock {Using One-Sided RDMA Reads to Build a Fast, CPU-Efficient Key-Value
  Store}.
\newblock In {\em USENIX ATC}, 2013.

\bibitem{timely_sigcomm_2015}
Radhika Mittal, Vinh~The Lam, Nandita Dukkipati, Emily Blem, Hassan Wassel,
  Monia Ghobadi, Amin Vahdat, Yaogong Wang, David Wetherall, and David Zats.
\newblock {TIMELY: RTT-Based Congestion Control for the Datacenter}.
\newblock In {\em SIGCOMM}, 2015.

\bibitem{irn_sigcomm_2018}
Radhika Mittal, Alexander Shpiner, Aurojit Panda, Eitan Zahavi, Arvind
  Krishnamurthy, Sylvia Ratnasamy, and Scott Shenker.
\newblock {Revisiting Network Support for RDMA}.
\newblock In {\em SIGCOMM}, 2018.

\bibitem{neugebauer2018pcie}
Rolf Neugebauer, Gianni Antichi, Jos\'{e}~Fernando Zazo, Yury Audzevich, Sergio
  L\'{o}pez-Buedo, and Andrew~W. Moore.
\newblock {Understanding PCIe Performance for End Host Networking}.
\newblock In {\em SIGCOMM}, 2018.

\bibitem{BO_github}
Fernando Nogueira.
\newblock {{Bayesian Optimization}: Open source constrained global optimization
  tool for {Python}}.
\newblock \url{https://github.com/fmfn/BayesianOptimization}, 2014.

\bibitem{storm2019systor}
Stanko Novakovic, Yizhou Shan, Aasheesh Kolli, Michael Cui, Yiying Zhang,
  Haggai Eran, Boris Pismenny, Liran Liss, Michael Wei, Dan Tsafrir, and Marcos
  Aguilera.
\newblock {Storm: A Fast Transactional Dataplane for Remote Data Structures}.
\newblock In {\em SYSTOR}, 2019.

\bibitem{osubench}
{OSU benchmarks}.
\newblock \url{https://mvapich.cse.ohio-state.edu/benchmarks/}, 2021.

\bibitem{perftest}
{OFED perftest}.
\newblock \url{https://github.com/linux-rdma/perftest}, 2021.

\bibitem{pfcrfc}
{IEEE DCB. 802.1Qbb - Priority-based Flow Control.}
\newblock \url{https://1.ieee802.org/dcb/802-1qbb/}, 2021.

\bibitem{hyv_2015_vee}
Jonas Pfefferle, Patrick Stuedi, Animesh Trivedi, Bernard Metzler, Ionnis
  Koltsidas, and Thomas~R. Gross.
\newblock {A Hybrid I/O Virtualization Framework for RDMA-Capable Network
  Interfaces}.
\newblock In {\em VEE}, 2015.

\bibitem{gfc2019sigcomm}
Kun Qian, Wenxue Cheng, Tong Zhang, and Fengyuan Ren.
\newblock {Gentle Flow Control: Avoiding Deadlock in Lossless Networks}.
\newblock In {\em SIGCOMM}, 2019.

\bibitem{rdma-core}
{Linux rdma-core}.
\newblock \url{https://github.com/linux-rdma/rdma-core}, 2021.

\bibitem{rdmaturing2021}
Waleed Reda, Marco Canini, Dejan Kostić, and Simon Peter.
\newblock {RDMA is Turing complete, we just did not know it yet!}, 2021.

\bibitem{schmilo2017graybox}
Sergej Schumilo, Cornelius Aschermann, Robert Gawlik, Sebastian Schinzel, and
  Thorsten Holz.
\newblock {kAFL: Hardware-Assisted Feedback Fuzzing for {OS} Kernels}.
\newblock In {\em USENIX Security}, 2017.

\bibitem{shi2016graph}
Jiaxin Shi, Youyang Yao, Rong Chen, Haibo Chen, and Feifei Li.
\newblock {Fast and Concurrent {RDF} Queries with RDMA-Based Distributed Graph
  Exploration}.
\newblock In {\em OSDI}, 2016.

\bibitem{1rma2020sigcomm}
Arjun Singhvi, Aditya Akella, Dan Gibson, Thomas~F. Wenisch, Monica Wong-Chan,
  Sean Clark, Milo M.~K. Martin, Moray McLaren, Prashant Chandra, Rob Cauble,
  Hassan M.~G. Wassel, Behnam Montazeri, Simon~L. Sabato, Joel Scherpelz, and
  Amin Vahdat.
\newblock {1RMA: Re-Envisioning Remote Memory Access for Multi-Tenant
  Datacenters}.
\newblock In {\em SIGCOMM}, 2020.

\bibitem{lite2017}
Shin-Yeh Tsai and Yiying Zhang.
\newblock {LITE Kernel RDMA Support for Datacenter Applications}.
\newblock In {\em SOSP}, 2017.

\bibitem{xue2019rdma}
Jilong Xue, Youshan Miao, Cheng Chen, Ming Wu, Lintao Zhang, and Lidong Zhou.
\newblock {Fast Distributed Deep Learning over RDMA}.
\newblock In {\em EuroSys}, 2019.

\bibitem{zhang2019rdma}
Yiwen Zhang, Yue Tan, Brent Stephens, and Mosharaf Chowdhury.
\newblock {Justitia: Software Multi-Tenancy in Hardware Kernel-Bypass
  Networks}.
\newblock In {\em NSDI}, 2022.

\bibitem{zhu2015dcqcn}
Yibo Zhu, Haggai Eran, Daniel Firestone, Chuanxiong Guo, Marina Lipshteyn,
  Yehonatan Liron, Jitendra Padhye, Shachar Raindel, Mohamad~Haj Yahia, and
  Ming Zhang.
\newblock {Congestion Control for Large-Scale RDMA Deployments}.
\newblock In {\em SIGCOMM}, 2015.

\end{thebibliography}

\appendix
\newpage

\section{Performance Anomalies Found}
\label{app:anomalies}
More details of these anomalies and the lesson we learn are included in this section. We present a concrete example of each anomaly and try our best to simplify each anomaly so that they can be reproduced easier. It is possible to find milder or stricter conditions that trigger the anomaly. We, to the best of our knowledge, also categorize these performance anomalies to their root causes based on our observation and conversations with our vendors. 
\subsection{Subsystem F with Mellanox 200\,Gbps CX-6 VPI}
\textbf{Root cause \#1: Receive WQE cache misses bottleneck RNIC receiving rate.}

\textit{(New) Anomaly \#1: UD with large WQE batch size and long WQ causes PFC pause frames and drastic throughput drop.}

Collie observes that the pause duration ratio can be up to $\approx$ 20.0\% with only a single UD QP. The pause duration ratio means that RNIC is asking the corresponding switch port to pause for $\approx$ 200 milliseconds within one second on average. We share the NIC vendor with our traffic engine tool and the running command. They have reproduced the anomaly in their environments, but the root cause is still not clear yet. Therefore, we claim this anomaly not fixed yet. To the best of our knowledge, it is likely due to the cache miss triggered by the pre-fetch mechanism for the receive WQE. This bottlenecks the receiver from receiving traffic. 

Here is a simplified concrete trigger setting of Anomaly \#1: There is 1 connection of UD QP using SEND/RECV Opcode. Each QP has 1 sending MR of 64KB and 1 receiving MR of 64KB. Each QP has a work queue of length 256 (i.e., \textit{max\_send/recv\_wr} = 256). The MTU is 2KB. The sender keeps sending 64 requests in a batch. Each request only has one SG element and a fixed size of 2KB.

\textit{(New) Anomaly \#2: UD with small WQE batch size, long WQ, small messages, and a few connections causes throughput to drop without pause frames.}

This anomaly is similar to \#1 but more tricky and has a different end-to-end symptom. Unlike \#1, Collies does not observe PFC pause frames when the setting is slightly different from \#1: if the sender does not post sending requests in batch or the batch size is small (e.g., less than 8) and the messages are relatively small (e.g., 512B, 1KB), the throughput will drop by more than 20\% without any PFC pause frame triggered when the receiver has an extremely long work queue. If we set a smaller work queue for the receiver, the throughput returns to the line rate. This anomaly is also reproduced and acknowledged by NIC vendor. We conjecture that it has a similar root cause to \#1, but due to unknown RNIC bottlenecks, it behaves differently that the throughput drops without pause frame.

Here is a simplified concrete trigger setting of Anomaly \#2: There are 16 connections of UD QP using SEND/RECV Opcode. Each QP has 1 sending MR of 64KB and 1 receiving MR of 64KB. Each QP has a work queue of length 1024. The MTU is 1KB. The sender keeps sending 4 requests in a batch. Each request only has one SG element of 1KB.

\textit{(New) Anomaly \#3: RC READ with large messages causes PFC pause frames when MTU is under 1500 (the default MTU for Ethernet).}

We observe the throughput drops drastically once we use RDMA READ opcode with 1500 MTU (1024 for RDMA), the default value for our data centers. The pause duration can be up to ~10\% and throughput drops to less than half. We report this to our NIC vendor and they tell us the low MTU may trigger the RNIC internal packet processing bottleneck for this 200\, Gbps NIC. We carefully survey the potential effect of MTU modification in our deployment and modify the MTU from 1500 to 4200, which supports 4096 as RDMA MTU. This anomaly is successfully fixed in this way.

Here is a simplified concrete trigger setting of Anomaly \#3: There are 8 connections of RC QP using Read opcode. Each QP has 1 sending MR of 4MB and 1 receiving MR of 4MB. Each QP has a work queue of length 128. The MTU is 1KB. The sender keeps sending RDMA READ requests. Each request only has one SG element and a fixed size of 4MB.

\textit{(New) Anomaly \#4: Bidirectional RC READ with large WQE batch size, long SG list, and a few connections causes PFC pause frames, even when MTU is set to 4200 (4096 for RDMA).}

This anomaly is tricky but severe. Even with 4200 MTU (Anomaly \#3 is solved), \sys observes about 30\% PFC pause duration ratio that when bidirectional RDMA READ happens and both sides post a large number of requests in a batch (e.g., 32), each request consists of multiple scatter gather element (e.g., 4) and there are a few connections (e.g., $\approx$ 160). As usual, this newly found anomaly is reported to the vendor and they have reproduced and confirmed the anomaly. For now, the root cause of this anomaly is still unknown. Therefore, we claim this anomaly not fixed yet.

Here is a simplified concrete trigger setting of Anomaly \#4: There are 80 connections of RC QP using Read opcode for each direction. Each QP has 1 sending MR of 64KB and 1 receiving MR of 64KB. Each QP has a work queue of length 128. The MTU is 4KB. The sender keeps sending 128 requests in a batch. Each request has 4 SG elements and a fixed size of 128B.

\textit{(New) Anomaly \#5: RC SEND with small MTU, large WQE batch, long WQ, and long messages causes PFC pause frames and drastic throughput drop.}

\textit{(New) Anomaly \#6: RC SEND with small MTU, small WQE batch, large SG list batch, long WQ, small messages, and a few connections causes reduced throughput without any pause frame.}

They are similar to UD ones (Anomaly \#1 and \#2) but have a more complex and stricter trigger. For example, \sys observes such anomaly only when MTU is small (e.g., 1024 for RDMA), work depth exceeds 1K for each QP as well as post multiple receive WQE in a batch. These anomalies are different because they have different QP types and stricter trigger conditions. For example, those anomalous application workloads in \#1 and \#2 won't trigger anomalies if we only switch the type of QP from UD to RC. Several discussion with our vendors tells us that the \textit{Reliable Connection} type contains some subtle variance inside the RNIC that result in such difference. These two are currently not fixed yet.

Here is a simplified concrete trigger setting of Anomaly \#5: There is 1 connection of RC QP using SEND/RECV opcode. Each QP has 1 sending MR of 64KB and 1 receiving MR of 64KB. Each QP has a work queue of length 1024. The MTU is 1KB. The sender keeps sending 64 requests in a batch. Each request has 2 SG elements and a fixed size of 2KB.

Here is a simplified concrete trigger setting of Anomaly \#6: There are 32 connections of RC QP using SEND/RECV opcode. Each QP has 1 sending MR of 64KB and 1 receiving MR of 64KB. Each QP has a work queue of length 1024. The MTU is 1KB. The sender keeps sending 8 requests in a batch. Each request has 2 SG elements and a fixed size of 1KB.

\textbf{Root cause \#2: Interconnect Context Memory cache misses reduce RNIC sending rates.}

\textit{(New) Anomaly \#7: RC WRITE with many QPs, small messages, small WQ depth, and small WQE batch size causes reduced throughput.}

\textit{(New) Anomaly \#8: RC WRITE with many MRs, small messages, and small WQE batch size causes reduced throughput.}

Though these two anomalies are well-known as the RDMA scalability problem, our real applications do not meet them even when the number of QPs exceeds 10K and the number of MRs exceeds 100K. However, \sys uncovers these two so we classified them into \textit{New} anomalies. We take a deep look into how \sys discovers them and have many discussions with our vendors. We find our experience interesting and worthy of sharing: RNIC caches many necessary structures on its cache (e.g., memory translation table and connection context). When a request triggers cache miss, the RNIC has to issue extra PCIe operation to fetch them from the host DRAM. This will certainly induce extra PCIe latency for processing this request (victim request). However, RNIC is highly pipelined, so even when the victim request has finished the PCIe operation, it may still have to wait for the other pipeline stages to get ready (e.g., a previous long egress request blocks this short egress request). Therefore, if the request size is relatively large enough, the cache miss will not have a large effect on end-to-end performance because the overhead is hidden due to the pipeline. 

Here is a simplified concrete trigger setting of Anomaly \#7: There are 480 connections of RC QP using RDMA WRITE opcode. Each QP has 1 sending MR of 64KB and 1 receiving MR of 64KB. Each QP has a work queue of length 16. The MTU is 1KB. The sender keeps sending requests without WQE batch. Each request has 1 SG element and a fixed size of 512B.

Here is a simplified concrete trigger setting of Anomaly \#8: There are 24 connections of RC QP using RDMA WRITE opcode. Each QP has 1024 sending MR of 64KB and 1024 receiving MR of 64KB. Each QP has a work queue of length 128. The MTU is 1KB. The sender keeps sending requests without WQE batching. Each request has 1 SG element and a fixed size of 512B.

\textbf{Root cause \#3: PCIe controller blocks RNIC from reading host memory.}

\textit{(Old) Anomaly \#9: Bidirectional traffic with a mixture of small and large messages in an SG list on particular AMD servers causes PFC pause frames and drastic throughput drop.}

This anomaly is found by one of our production applications that keeps sending such message patterns (described in \ref{sec:bg}). The root cause of this anomaly is due to PCIe ordering issue. If the RNIC on the AMD server is not configured as PCIe relaxed ordering device, a DMA request may be blocked by the previous one. Therefore, when bidirectional traffic with a mix of short and long requests. The ingress short requests, together with the completion of egress traffic, blocks the ingress long requests. This results in RNIC buffer accumulation and triggers a large amount of PFC pause frames. The throughput can only achieve ~60\,Gbps with 25\% pause frame duration ratio on average. With much effort from our appreciative vendors, we finally fix this by configuring RNIC as a forced relaxed ordering PCIe device.  

Here is a simplified concrete trigger setting of Anomaly \#9: There are 8 connections of RC QP using RDMA WRITE opcode for each direction. Each QP has 1 sending MR of 4MB and 1 receiving MR of 4MB. Each QP has a work queue of length 128. The MTU is 4KB. The sender keeps sending 8 requests in a batch. Each request has 3 SG elements and the pattern is [128B, 64KB, 1KB].

\textbf{Root cause \#4: RNIC packet processing bottleneck.}

\textit{(New) Anomaly \#10: Bidirectional RC Write with large WQE batch size, a mixture of long messages and lots of short messages, and a few connections causes PFC pause frames.}  

\sys finds that when several RC QPs keep posting multiple short requests (e.g., 64B, 128B) in batch and a few long requests for both directions, a large amount of pause duration is triggered. This RNIC of the RDMA subsystem has already been configured as forced relaxed ordering PCIe device (Anomaly \#8 is solved). Our vendors have confirmed this anomaly and announce it fixed in their upcoming firmware release. The lengthy discussion with our vendor shows us the rough root cause: some component for packet processing inside the RNIC is not fully bidirectional, and our bidirectional reliable traffic (requires packet-level ACK) pattern with a huge amount of short requests, trigger that component's bottleneck. This results in long requests blocked and then many PFC pause frames are generated. 

Here is a simplified concrete trigger setting of Anomaly \#10: There are 320 connections of RC QP using RDMA WRITE opcode for each direction. Each QP has 1 sending MR of 64KB and 1 receiving MR of 64KB. Each QP has a work queue of length 128. The MTU is 1KB. The sender keeps sending 64 requests in a batch. Each request has 1 SG element and the pattern is [64KB, 128B, 128B, 128B].

\textbf{Root cause \#5: Host topology causes PCIe latency to increase, and this bottlenecks RNIC receiving rate.} 

\textit{(New) Anomaly \#11: On specific types of AMD servers, Bidirectional cross-socket traffic causes pause frame storm and drastic throughput drop.}

\sys outputs the minimal feature set with only source/destination NUMA set and bidirectional traffic, indicating these two are the dominant factors. With this bidirectional (A to B and B to A) cross-socket NUMA setting (e.g., NUMA 0 from socket 0 for A and NUMA 2 from socket 1 for B, where socket 0 is the affinitive node for RNIC), even mild traffic with only a single connection can trigger up to 15.7\% pause frame duration ratio.
After several conversations with our RNIC and server vendors, we conjecture the root cause lies in these particular servers' cross-socket performance because we run the same traffic with the same NIC on different servers but do not observe the same phenomenon. We consider this anomaly as fixed because the vendor helps us roughly understand the root cause and suggest we use 2x100\,Gbps NIC (each for a socket) to reduce cross-socket traffic, and we follow this guidance.

Here is a simplified concrete trigger setting of Anomaly \#11: There is 1 connection of RC QP using RDMA WRITE opcode for each direction. Each QP has 32 sending MR of 4MB and 32 receiving MR of 4MB. Each QP has a work queue of length 128. The MTU is 4KB. The sender keeps sending 16 requests in a batch. Each request has 1 SG element with a fixed size of 256KB. The QP on host A is using the memory of socket 0 and the QP on host B is using the memory of socket 1.

\textit{(Old) Anomaly \#12: GPU-direct RDMA causes pause frame storm and drastic throughput drop on particular AMD servers.} 

We observe a huge amount of pause frames and drastic throughput drop only on some servers in our clusters. The pause duration ratio can be up to ~15\% and throughput can drop to less than 20\% (i.e., 40\,Gbps) in this scenario. After careful debugging with our NIC vendor's strong support, we find out that there is a slight difference in PCIe bridge configuration (PCIe ACSCtl) between the anomalous server and normal ones. The anomalous configuration will forward GPU traffic to the root complex rather than directly to the RNIC. We fix this anomaly by adopting the correct configuration.

Here is a simplified concrete trigger setting of Anomaly \#12: There are 8 connections of RC QP using RDMA WRITE opcode for each direction. Each QP has 1 sending MR of 4MB and 1 receiving MR of 4MB. Each QP has a work queue of length 128. The MTU is 4KB. The sender keeps sending 8 requests in a batch. Each request has 3 SG elements and the pattern is [128B, 64KB, 1KB]. All MRs are allocated from GPU memory and we use the GPU under the same PCIe bridge (i.e., shown as PIX/PXB in \textit{nvidia-smi} result).

\textbf{Root cause \#6: RDMA NIC has potential in-NIC incast/congestion.} 

\textit{(Old) Anomaly \#13: Co-existence of receiving traffic and loopback traffic causes PFC pause frames.} 

This anomaly is found in our real applications and can also be uncovered by Collie. Our machine learning system runs workers and servers, and they use RDMA to accelerate the communication. However, once a worker and a server are scheduled on the same physical machine, there will be loopback traffic: the worker will send RDMA traffic to the server on the same host. Meanwhile, the server is receiving traffic from workers on other physical machines. The combination of receiving and loopback traffic triggers congestion/incast inside the NIC. And this RNIC lacks a mechanism to limit the loopback traffic rate, which makes the problem worse. After several discussions with our vendor, we bypass this anomaly by identifying the loopback communication and using other IPC mechanisms (e.g., shared memory). We do not consider this anomaly fixed because we cannot fully rely on other IPC mechanisms, especially for the virtualization environment. This anomaly exposes that a proper design of RNIC needs to consider NIC incast and we are glad to see that some latest RNIC have done so.

Here is a simplified concrete trigger setting of Anomaly \#13: There are 16 connections of RC QP using RDMA WRITE opcode. 16 receivers are 8 senders are on the same host A and the other 8 senders are on the host B. Each QP has 32 sending MR of 4MB and 32 receiving MR of 4MB. Each QP has a work queue of length 128. The MTU is 4KB. The sender keeps sending 16 requests in a batch. Each request has 1 SG element with a fixed size of 256KB. 

\subsection{Subsystem H with Broadcom 100\,Gbps P2100G}
\textit{(New) Anomaly \#14: Bidirectional RC traffic with lots of connections and the large MTU causes reduced throughput without PFC pause frame.}

Collie observes that a large MTU is necessary to trigger this anomaly. Once we switch the MTU from 4096 (for RDMA) to 1024, both directions can achieve the line rate. This is unusual because most cases show that large MTU improves the performance and small MTU triggers performance anomalies. We don't observe the same phenomenon on any other type of RNICs.

Here is a simplified concrete trigger setting of Anomaly \#14: There are 1024 connections of RC QP using RDMA WRITE opcode for each direction. Each QP has 81 sending MR of 256KB and 83 receiving MR of 256KB. Each QP has a work queue of length 128. The MTU is 4KB. The sender keeps sending 1 request in a batch. Each request has 4 SG element with a fixed size of 64KB.

\textit{(New) Anomaly \#15: UD with long WQ and lots of connections causes PFC pause frames.}

This anomaly is similar to the Mellanox anomaly \#1 but has a slightly different trigger. \sys successfully trigger \#1 with only a single connection, but for P2100 RNIC our multiple runs show that a few connections are necessary.

Here is a simplified concrete trigger setting of Anomaly \#15: There are 32 connections of UD QP using SEND/RECV opcode. Each QP has 1 sending MR of 4KB and 1 receiving MR of 4KB. Each QP has a work queue of length 64. The MTU is 2KB. The sender keeps sending 1 request in a batch. Each request has 1 SG element. The message pattern is like [256B, 1KB, 64B, 1KB].

\textit{(New) Anomaly \#16: RC READ with lots of connections, large WQE batch size, and small MTU causes PFC pause frames.}

This anomaly is similar to the Mellanox anomaly \#4 and it shows that for the same RNIC and other hardware components, the best MTU choice can be different when workloads change.

Here is a simplified concrete trigger setting of Anomaly \#16: There are 500 connections of RC QP using RDMA READ opcode. Each QP has 1 sending MR of 256KB and 1 receiving MR of 256KB. Each QP has a work queue of length 128. The MTU is 1KB. The sender keeps sending 8 requests in a batch. Each request has 1 SG element with a fixed size of 64KB.

\textit{(New) Anomaly \#17: RC SEND with lots of connections, small WQE batch size, small MTU, short messages, and long WQ causes PFC pause frames.}

We have reported this anomaly to our vendor. To the best of our knowledge, we conjecture this anomaly is related to some corresponding WQE cache component inside RNIC. 

Here is a simplified concrete trigger setting of Anomaly \#17: There are 80 connections of RC QP using SEND/RECV opcode. Each QP has 1 sending MR of 1MB and 1 receiving MR of 1MB. Each QP has a work queue of length 128. The MTU is 1KB. The sender keeps sending 1 request per batch. Each request has 1 SG element of fixed size 1KB. 

\textit{(New) Anomaly \#18: Bidirectional RC WRITE with a few connections, large WQE batch, and small messages causes PFC pause frames.}

Our vendor has confirmed anomalies \#17 and \#18. They have reproduced these two anomalies and help us fix them. The solution is to configure some specific registers of the RNIC, and these two anomalies disappear. 

Here is a simplified concrete trigger setting of Anomaly \#18: There are 16 connections of RC QP using RDMA WRITE for each direction. Each QP has 1 sending MR of 12KB and 1 receiving MR of 12KB. Each QP has a work queue of length 64. The MTU is 1KB. The sender keeps sending 16 requests in a batch. Each request has 1 SG element of fixed size 64KB.

\end{document}